\documentclass[11pt]{article}
\usepackage{amsmath,amssymb,graphicx}
\usepackage{hyperref}
\usepackage[nosort,compress]{cite}
\usepackage{ytableau}
\usepackage{tikz}
\usetikzlibrary{tikzmark}
\usepackage{multicol,color,xcolor,longtable}
\usepackage[title]{appendix}

\definecolor{darkred}{rgb}{0.65,0.15,0}
\hypersetup{pdfborder={0 0 0},colorlinks=true,urlcolor=blue,citecolor=blue,linkcolor=darkred,linktocpage=true}

\newcommand{\nn}{\nonumber}

\newcommand{\ALT}{\textrm{\large{$\wedge$}}}

\newcommand{\mf}[1]{{\mathfrak{#1}}}

\newcommand{\be}{\begin{equation}}
\newcommand{\ee}{\end{equation}}
\newcommand{\bea}{\begin{eqnarray}}
\newcommand{\eea}{\end{eqnarray}}

\makeatletter

\@addtoreset{equation}{section}
\makeatother

\begin{document}

\hypersetup{pageanchor=false}
\mbox{}
\vspace{15mm}

\begin{center}
	{\LARGE \bf \sc Non-Lorentzian expansions of the Lorentz force\\[5mm] and kinematical algebras}\\[20mm]

José Luis V. Cerdeira$ {}^1 $,
Joaquim Gomis${}^1$ and Axel Kleinschmidt${}^{2,3}$ \\[3mm]
\footnotemark[1]{\it Departament de F\'isica Qu\`antica i Astrof\'isica\\
 and Institut de Ci\`encies del Cosmos (ICCUB), Universitat de Barcelona\\
Mart\'i i Franqu\`es , 08028 Barcelona, Spain}\\[1mm]
\footnotemark[2]{\it  Max-Planck-Institut f\"ur Gravitationsphysik\\
     Albert-Einstein-Institut \\
     Am M\"uhlenberg 1, 14476 Potsdam, Germany}\\[1mm]
\footnotemark[3]{\it International Solvay Institutes\\
ULB-Campus Plaine CP231, 1050 Brussels, Belgium}\\[25mm]

\begin{abstract}
\noindent
	We consider non-Lorentzian expansions, Galilean and Carrollian, of the Lorentz force equation in which both the particle position and the electro-magnetic field are expanded. There are two well-known limits in the case of a constant field, called electric and magnetic, that are studied separately. We show that the resulting equations of motion follow equivalently from considering a non-linear realisation of a certain infinite-dimensional algebras.

\end{abstract}
\end{center}

\thispagestyle{empty}

\newpage
\hypersetup{pageanchor=true}
\setcounter{page}{1}

\setcounter{tocdepth}{2}
\tableofcontents

\section{Introduction}
	
	Non-Lorentzian theories refer to theories that have as their underlying symmetry algebra a kinematical algebra that is different from the Poincaré one, such as the Galilei algebra or the Carroll algebra, for a review see \cite{Bergshoeff:2022eog}. There is a variety of kinematical algebras that have been classified in~\cite{Bacry:1968zf,Bacry:1986,Figueroa-OFarrill:2017sfs,Figueroa-OFarrill:2017ycu,Figueroa-OFarrill:2018ilb}---in this paper we focus on the Galilean and Carrollian case. Usually, non-Lorentzian systems can be obtained as the limit of a relativistic system when some characteristic parameter  goes to zero (infinity). Consider for example a relativistic free point particle and its velocity relative to the speed of light $v/c$. Taking this parameter to zero (infinity) one obtains the Galilean (Carrollian) free particle.
Non-relativistic expansions in $1/c^2$ have been considered in various previous studies, see for example the review~\cite{Hartong:2022lsy} in the context of gravity.

Given a relativistic system, instead of considering its strict non-Lorentzian limits, one can perform a non-Lorentzian expansion in terms of the characteristic parameter, that allows one to obtain not only the non-Lorentzian limit, but also a series of corrections. However, only the first term in the expansion exhibits the symmetry of the contracted (non-Lorentzian) algebra, whereas the full expansion exhibits the relativistic symmetry. In \cite{Gomis:2022spp} it was shown how to study the symmetry algebra of the truncated expansions at any level. The idea is to construct, from the contracted algebra, $\mathfrak{g}_0:=\mathfrak{g}^{(0)}$  with generators $\{t^{(0)}_\alpha\}$ an infinite sequence of expansions $\mathfrak{g}^{(N)}$ with generators $\{t_\alpha^{(n)}\}_{0\leq n\leq N}$, leading to an infinite-dimensional Lie algebra $\mathfrak{g}_{\infty}$. 
The method of Lie algebra expansions was pioneered in~\cite{Hatsuda:2001pp,Boulanger:2002bt,deAzcarraga:2002xi,Izaurieta:2006zz,deAzcarraga:2007et,Bergshoeff:2019ctr}.
The infinite-dimensional algebra $\mathfrak{g}_{\infty}$ is like a non-Lorentzian
expansion of the contracted algebra. Since $\mathfrak{g}$ acts on the space-time manifold $M$, its infinite expansion $\mathfrak{g}_{\infty}$ acts on an infinite-dimensional homogeneous space $M^{(\infty)}$ using non-linear realisations. Introducing \textit{collective coordinates} on this generalised space, one can recover the space $M$ and the symmetry algebra $\mathfrak{g}$. 

In \cite{Gomis:2019sqv} it was shown that, starting from the Poincar\'e algebra, 
 one could obtain a non-relativistic expansion of the relativistic free particle Lagrangian $\mathcal{L}= -mc \sqrt{-\dot{x}^2}$, by considering an infinite-dimensional algebra $\mathfrak{G}_{\infty}$ containing the Galilei algebra $\mathfrak{G}$ as a quotient by an ideal. The non-linear realisation of $\mathfrak{G}_{\infty}$ in a top-down approach using an associated infinite-dimensional space-time yields associated Euler--Lagrange equations that correspond to the non-relativistic expansion of equations of motion of a free relativistic particle if we consider a specific slice of $M^{(\infty)}$.

In this work, for the case of a particle in a constant electro-magnetic field,
we will start considering the inverse procedure of that in the free case. Starting from the bottom up,
we will consider a non-relativistic expansion of the relativistic Lorentz force equation. Both the particle position and the electro-magnetic field are expanded.

 In \cite{Schrader:1972zd}
 it was shown that the Poincaré algebra admits a non-central extension, the {Maxwell} algebra. A standard Lagrangian which realises this symmetry algebra is \cite{Bonanos:2008ez}\footnote{Compared to~\cite{Gomis:2017cmt} we have changed the sign of the unit charge of the particle, that is the sign in front of the $f_{\mu\nu}$ term.}
\begin{align}
 \mathcal{L}=-mc\sqrt{-\dot{x}^\mu \dot{x}_\mu}-\frac 12 f_{\mu\nu}\Omega^{\mu\nu}
 \end{align}
 where the dot denotes $d/d\tau$ and $\Omega^{\mu\nu}:=\dot\theta^{\mu\nu}+\frac 12\left( \dot{x}^\mu x^\nu-\dot{x}^\nu x^\mu\right) $ is a Maurer--Cartan derivative, while $f_{\mu\nu}(\tau)$ and $\theta^{\mu\nu}(\tau)$ are new dynamical variables, that are similar to higher inertial moments \cite{Dixon:1967,Gomis:2017cmt}. This Lagrangian describes a 
relativistic particle subject to a generic external, constant electro-magnetic field represented by $f_{\mu\nu}$. 

There is more than one  non-Lorentzian limit for electro-magnetism~\cite{LeBellac:1973,Duval:2014uoa,Barducci:2019fjc,Gomis:2019fdh}. Different regimes arise depending on the relative strength of the magnetic and electric field. 
The limits are called electric and magnetic depending on the dominant component of the electro-magnetic field.\footnote{There is a third limit, called pulse in~\cite{Barducci:2019fjc,Gomis:2019fdh}, where the electric and magnetic field have similar strength. We will comment more on this limit in Section~\ref{sec:NRLor} and Appendix~\ref{app:1c}.}
We will see in both cases that they admit different infinite-dimensional symmetry algebras. 

As we shall show these infinite-dimensional algebras coincide with those obtained by applying the Lie algebra expansion method to the relativistic {Maxwell} algebra in two different ways. For instance in the Galilei case, these two ways differ by connecting to the non-relativistic Galilean electric $\mathfrak{E}$ and Galilean magnetic $\mathfrak{M}$ Maxwell algebras. The corresponding infinite-dimensional algebras $\mathfrak{E}_\infty$ and $\mathfrak{M}_\infty$ and their relation to the Galilean free algebras \cite{Gomis:2019fdh} will be exhibited.\footnote{Carrollian free algebras were discussed in~\cite{Gomis:2022spp}.} These infinite-dimensional algebras admit quotients that describe the symmetries of the expansion up to a finite order in $1/c$.
Using the non-linear realisation approach we construct Lagrangians associated with these infinite-dimensional algebras, whose Lagrange equation of motion coincide with the ones obtained by expanding the Lorentz equation in the bottom-up approach. The non-linear realisation also involves extra coordinates of the infinite-dimensional space-time and that are expanded.
The Lagrangian also involves extra fields on the particle's world-line that are related to the expansion of the electro-magnetic field.
We shall perform a similar analysis for the Carroll limit.

The organisation of this work is as follows: 
	In Section~\ref{sec:NRLor}, we will obtain a non-relativistic expansion of the Lorentz equation in powers of $1/c^2$ in the two cases depending on whether the constant electric or magnetic field is dominant.
	In Section~\ref{sec:topdown}, we will study the same problem, that of a non-relativistic expansion of the Lorentz equation, through its algebra of symmetries, the {Maxwell} algebra.
	The Carroll case, both bottom-up and top-down, is considered in Section~\ref{sec:Carroll}. 
	In Section~\ref{sec:duality}, we discuss the relation between the Galilei and the Carroll limit with electro-magnetic field.
	In appendices we also relate our algebraic constructions to that of free Lie algebras, as well as the case of a $1/c$ expansion rather than $1/c^2$.
	
\section{Non-relativistic expansion of the Lorentz equation}
\label{sec:NRLor}

Our starting point is the Lorentz equation for a massive test particle of unit charge in an electro-magnetic field background $ F^{\mu\nu} $
\begin{align}
	\label{Lorentz}
	mc \frac{d}{d\tau} \left( \frac{\dot{x}^\mu(\tau)}{\sqrt{-\dot{x}^2}} \right) =  F^{\mu\nu} (x(\tau)) \dot{x}_\nu(\tau)
\end{align}
where the dot denotes derivative with respect to the arbitrary worldline parameter $ \tau $, $ x^\mu $ is the position four-vector and we are using the metric $ \eta^{\mu\nu}=\text{diag}(-1,1,\dots,1) $, such that $\dot{x}^2 = \dot{x}^\mu \dot{x}_\mu = - (\dot{x}^0)^2 + \dot{\vec{x}}^2$. Greek indices $ \mu,\nu $ refer to space-time indices, while $ i,j,k,\dots $ indices will be reserved for the spatial components. In this way we write $x^\mu=(x^0,x^i)= (ct, x^i)$. 
From now on we assume that $F^{\mu\nu}$ is constant but not fixed. This means that under Lorentz transformations we have $x^\mu \to \Lambda^\mu{}_\nu x^\nu$ and $F^{\mu\nu} \to \Lambda^\mu{}_\rho \Lambda^\nu{}_\sigma F^{\rho\sigma}$. 
We think of the constant $F_{\mu\nu}$ as parametrising a moduli space of theories and the Lorentz transformation above transforms points on this moduli space.
If one were to fix the point on moduli space by taking a fixed $F_{\mu\nu}$, the Lorentz symmetry is broken from six to two generators called the Bacry--Combe--Richards algebra~\cite{Bacry:1970ye}.

Separating space and time, the Lorentz force  reads in non-manifestly covariant form
\begin{subequations}
\label{eq:LFsplit}
\begin{align}
\label{eq:LorT}
	m \frac{d}{d\tau} \left( \frac{1}{\sqrt{1- \dot{\vec{x}}^2 /( c\dot{t})^2}} \right) &= \frac1{c^2} \tilde{F}^{ti} \dot{x}_i \,,\\
\label{eq:LorS}
	m \frac{d}{d\tau} \left( \frac{\dot{x}^i}{\dot{t} \sqrt{1- \dot{\vec{x}}^2 /( c\dot{t})^2}} \right) &=  \tilde{F}^{ti} \dot{t} + F^{ij} \dot{x}_j\,,
\end{align}
\end{subequations}
where we have defined the rescaled field $ \tilde{F}^{ti}=cF^{0i} $ which has the same units as the electric field.

We expand the $ \gamma $ factor $ \gamma=1+\frac 12 \frac{\dot{\vec{x}}^2}{\dot{t}^2c^2}+\frac 3 8 \frac {\dot{\vec{x}}^4}{\dot{{t}}^4c^4}+\dots $ and construct an expansion of the coordinates $ x^\mu $ according to\footnote{Here, we are assuming an expansion in powers of $1/c^2$. In Appendix~\ref{app:1c}, we consider a $1/c$ expansion which is more appropriate for a certain limit of the Maxwell equations.} 
\begin{align}
	t=t_{(0)}+\frac{1}{c^2}t_{(1)}+\dots, && x^i=x^i_{(0)}+\frac 1 {c^2}x^i_{(1)}+\dots
\end{align}
as well as of the electro-magnetic field $ F^{\mu\nu} $ by
\begin{align}
	\label{Fexpansion}
	\tilde{F}^{ti}=\tilde{F}^{ti}_{(0)}+\frac 1 {c^2} \tilde{F}^{ti}_{(1)}+\dots && F^{ij}=F^{ij}_{(0)}+\frac 1 {c^2}F^{ij}_{(1)}+\dots
\end{align}

Order by order in $ 1/c^2 $, the following non relativistic expansion of the Lorentz equation~\eqref{eq:LFsplit} is obtained:
\begin{subequations}
\label{eq:expLF}
\begin{align}
	\label{expansion1}
	m \frac{d}{d\tau} \left[ \frac{\dot{\vec{x}}_{(0)}^2}{2\dot{t}_{(0)}^2} \right] &= \tilde{F}^{ti}_{(0)} \dot{x}_{(0) i}\\
		\label{expansion1b}
	m \frac{d}{d\tau} \left[  \frac{\dot{x}_{(0)}^i}{\dot{t}_{(0)}} \right] &= - \tilde{F}^{it}_{(0)}  \dot{t}_{(0)} + F^{ij}_{(0)} \dot{x}_{(0)j}\\
		\label{expansion1c}
	m  \frac{d}{d\tau}\left[ \frac{3}{8} \frac{ (\dot{\vec{x}}_{(0)}^2)^2}{\dot{t}_{(0)}^4} + \frac{\dot{\vec{x}}_{(0)} \cdot \dot{\vec{x}}_{(1)}}{\dot{t}_{(0)}^2}  - \frac{\dot{\vec{x}}_{(0)}^2 \dot{t}_{(1)}}{\dot{t}_{(0)}^3}
	\right] &=   \tilde{F}^{ti}_{(0)} \dot{x}_{(1) i} + \tilde{F}^{ti}_{(1)} \dot{x}_{(0) i} \\
	m  \frac{d}{d\tau}\left[  \frac12 \frac{\dot{\vec{x}}_{(0)}^2 \dot{x}_{(0)}^i }{ \dot{t}_{(0)}^3} - \frac{\dot{t}_{(1)} \dot{x}_{(0)}^i}{\dot{t}_{(0)}^2} + \frac{\dot{x}_{(1)}^i}{\dot{t}_{(0)}}   \right] 
	&=
	- \tilde{F}^{it}_{(0)}\dot{t}_{(1)}+  F^{ij}_{(0)}  \dot{x}_{(1)j}  - \tilde{F}^{it}_{(1)}\dot{t}_{(0)}+  F^{ij}_{(1)}  \dot{x}_{(0)j} \label{expansion2}\\
	\dots &\nn
\end{align}
\end{subequations}
Continuing these equations to infinite order is still fully equivalent to the relativistic system~\eqref{eq:LFsplit}.
The equations~\eqref{expansion1} and~\eqref{expansion1b} are the standard non-relativistic particle in a general electro-magnetic field. In particular,~\eqref{expansion1} is the conservation of energy.

We are now interested in non-Lorentzian limits of the charged particle. 
It is known that there is not a single non-relativistic limit for electro-magnetism~\cite{LeBellac:1973,Duval:2014uoa,Barducci:2019fjc,Gomis:2019fdh}. Instead, different regimes appear depending on the relative strength of the magnetic and electric field. In the \textit{magnetic limit} $ |\mathbf{E}|/|\mathbf{B}|\ll c $ whereas in the \textit{electric limit}  $|\mathbf{E}|/|\mathbf{B}|\gg c $.\footnote{There is a third limit, called pulse in~\cite{Barducci:2019fjc,Gomis:2019fdh}, where the electric and magnetic field have similar strength, i.e. $ |\mathbf{E}|=c|\mathbf{B}| $. This limit is not compatible with the expansion proposed in (\ref{Fexpansion}) but instead requires a $1/c$ expansion that will be treated in Appendix~\ref{app:1c}.}
Since $|{\bf E}|^2 - c^2 |{\bf B}|^2$ is Lorentz-invariant, the notion of the different limits is independent of the choice of frame.

\subsection{Magnetic limit}

The magnetic limit can be obtained by having $ F^{ij}_{(0)}\neq 0 $ and  keeping $ \tilde{F}^{ti} $ fixed, since then
\begin{align}
	F^{ij}\gg \frac 1 c \tilde{F}^{ti}
\end{align}
is satisfied. The particle's equations in the magnetic limit are therefore formally the same as before taking the limit, namely~\eqref{eq:expLF}. However, the limit in the Maxwell field equations in the magnetic case are only non-relativistically invariant and lack the displacement current~\cite{LeBellac:1973,Duval:2014uoa}.

The equations~\eqref{eq:expLF} are invariant under the following transformation laws:
\begin{subequations}
\label{eq:trmmag}
\begin{align}
	\delta t_{(n)}&=\epsilon_{(n)}+\sum_{m=0}^{n-1} \vec{v}_{(m)}\cdot \vec{x}_{(n-m-1)},&&	\delta x^i_{(n)}=\epsilon^i_{(n)}+\sum_{m=0}^n v^i_{(m)}t_{(n-m)}\,, \\
	\delta \tilde{F}^{ti}_{(n)}&= \sum_{m=0}^n v_{(m)k}F^{ki}_{(n-m)}, && \delta F^{ij}_{(n)} = -2\sum_{m=0}^{n-1} \tilde{F}^{t[i}_{(m)}v^{j]}_{(n-m-1)}\,,&
\end{align}
\end{subequations}
where we have introduced the expansion of the boost parameter $ v^i=v^i_{(0)}+\frac 1 {c^2} v^i_{(1)} +\dots$ and also expanded time and spatial translations, $ \epsilon=\epsilon_{(0)}+\frac 1 {c^2} \epsilon_{(1)}+\dots $ and $ \epsilon^i=\epsilon^i_{(0)}+\frac 1 {c^2} \epsilon^i_{(1)}+\dots $

These transformations can be obtained as the non-relativistic expansion of the infinitesimal Lorentz transformations.
The transformations for the electro-magnetic fields are the Galilean expansion of the Lorentz transformation of $F_{\mu\nu}$ relating different constant fields, thus moving on moduli space.
As we shall show in the next section the transformations laws  for $ t_{(n)} $ and $ x^i_{(n)} $ are the same as for the action of the $ \mathfrak{G}_\infty $ algebra that was introduced in~\cite{Gomis:2019sqv}.
The transformations for the electro-magnetic fields will also be seen to agree with those derived from an extended algebra.

The boost operator defines a sequence:
\begin{equation}
\dots x^i_{(1)}\xrightarrow{v_{(1)}^i}t_{(1)}\xrightarrow{v_{(0)}^i}x^i_{(0)}\xrightarrow{v_{(0)}^i}t_{(0)}\xrightarrow{v_{(0)}^i}0\,,
\end{equation}
which extends the known two-step nilpotency $ \vec{x}\xrightarrow{\vec{v}}t \xrightarrow{\vec{v}}0$ of the Galilei algebra (see e.g. \cite{Gomis:2022spp}) to higher orders.

The sequence of boosts can also be used to relate the equations of motion in an indecomposable manner:
\begin{equation}
\dots \eqref{expansion2} \xrightarrow{v_{(1)}^i}\eqref{expansion1c}\xrightarrow{v_{(0)}^i}\eqref{expansion1b}\xrightarrow{v_{(0)}^i}\eqref{expansion1}\xrightarrow{v_{(0)}^i}0\,.
\end{equation}
This action is indecomposable as there is no transformation going in the other direction.

\subsection{Electric limit}

To obtain the electric limit, one can set $F^{ij}_{(0)}=0$, so that  for $\tilde{F}_{(0)}^{ti} \neq0$ one has
\begin{align}
	F^{ij}\ll \frac 1 c \tilde{F}^{ti}.
\end{align}
This leads to the following limit of the equations of motions~\eqref{eq:expLF}:
{\allowdisplaybreaks
\begin{subequations}
\label{eq:LFelec}
\begin{align}
	\label{perele1}
	m \frac{d}{d\tau} \left[ \frac{\dot{\vec{x}}_{(0)}^2}{2\dot{t}_{(0)}^2} \right] &= \tilde{F}^{ti}_{(0)} \dot{x}_{(0) i}\,,\\
	m \frac{d}{d\tau} \left[  \frac{\dot{x}_{(0)}^i}{\dot{t}_{(0)}} \right] &=  \tilde{F}^{ti}_{(0)}  \dot{t}_{(0)} \,,\\
	m  \frac{d}{d\tau}\left[ \frac{3}{8} \frac{ (\dot{\vec{x}}_{(0)}^2)^2}{\dot{t}_{(0)}^4} + \frac{\dot{\vec{x}}_{(0)} \cdot \dot{\vec{x}}_{(1)}}{\dot{t}_{(0)}^2}  - \frac{\dot{\vec{x}}_{(0)}^2 \dot{t}_{(1)}}{\dot{t}_{(0)}^3}
	\right] &=   \tilde{F}^{ti}_{(0)} \dot{x}_{(1) i} + \tilde{F}^{ti}_{(1)} \dot{x}_{(0) i} \,,\\
	m  \frac{d}{d\tau}\left[  \frac12 \frac{\dot{\vec{x}}_{(0)}^2 \dot{x}_{(0)}^i }{ \dot{t}_{(0)}^3} - \frac{\dot{t}_{(1)} \dot{x}_{(0)}^i}{\dot{t}_{(0)}^2} + \frac{\dot{x}_{(1)}^i}{\dot{t}_{(0)}}   \right] 
	&=
	\tilde{F}^{ti}_{(0)}\dot{t}_{(1)} +\tilde{F}^{ti}_{(1)}\dot{t}_{(0)}+  F^{ij}_{(1)}  \dot{x}_{(0)j} \,.
	\label{perele2}
\end{align}
\end{subequations}
These equations  are invariant instead under the transformations
\begin{subequations}
\label{eq:trmelec}
\begin{align}
	\delta t_{(n)}&=\epsilon_{(n)}+\sum_{m=0}^{n-1} \vec{v}_{(m)}\cdot \vec{x}_{(n-m-1)}\,,&&	\delta x^i_{(n)}=\epsilon^i_{(n)}+\sum_{m=0}^n v^i_{(m)}t_{(n-m)}\,,& \\
	\delta \tilde{F}^{ti}_{(n)}&= \sum_{m=0}^{n-1} v_{(m)k}F^{ki}_{(n-m-1)}\,, && \delta F^{ij}_{(n)} =\sum_{m=0}^{n} -2\tilde{F}^{t[i}_{(m)}v^{j]}_{(n-m)}&
\end{align}
\end{subequations}
that differ from the magnetic transformations~\eqref{eq:trmmag}. In the next section, we shall show how these equations and transformations obtained from a bottom-up approach can also be derived from the top down by using a suitable infinite-dimensional Lie algebra.}

\section{Lie-algebraic reformulation}
\label{sec:topdown}

A particle subject to a constant electro-magnetic field has symmetries extending the Poincar\'e algebra~\cite{Bacry:1970ye}. If one transforms also the electro-magnetic background under Lorentz transformations, the associated algebra of symmetries is the {Maxwell} algebra~\cite{Schrader:1972zd,Bonanos:2008ez,Gomis:2017cmt} which is a non-central extension of the {Poincar\'e} algebra, with generators $ P_\mu,M_{\mu\nu} $ and $ Z_{\mu\nu} $. The algebra is given by
\begin{subequations}
\label{eq:Max}
\begin{align}
	{\left[M_{\mu\nu }, M_{\rho \sigma}\right] } &=\eta_{\nu\rho} M_{\mu \sigma}-\eta_{\nu\sigma} M_{\mu\rho}-\eta_{\mu\rho} M_{\nu\sigma}+\eta_{\mu\sigma } M_{\nu\rho}\,, \\
	{\left[M_{\mu\nu}, P_{\rho}\right] } &=\eta_{\nu\rho} P_{\mu}-\eta_{\nu\rho} P_{\mu} \,,\\
	{\left[P_{\mu}, P_{\nu}\right] } &=Z_{\mu\nu}\,.
\end{align} 
\end{subequations}
 The most general reparametrisation-invariant Lagrangian at first order in derivatives one could write realising this symmetry algebra is 
\begin{align}
\label{eq:MaxL}
\mathcal{L} =- mc\sqrt{-\dot{x}^\mu\dot{x}_\mu} -\frac 12 f_{\mu\nu}\Omega^{\mu\nu}\,,
\end{align}
 where $ \Omega^{\mu\nu}:=\dot\theta^{\mu\nu} +\frac 12 \left(\dot{x}^\mu x^\nu-\dot{x}^\nu x^\mu\right) $ is the Maurer--Cartan derivative and $ \theta^{\mu\nu} $ are the coordinates associated to the new generators $ Z_{\mu\nu} $. This Lagrangian describes a point particle subject to an external constant electro-magnetic field. The field $f_{\mu\nu}=f_{\mu\nu}(\tau)$ is set to a constant by the equations of motion of $\theta_{\mu\nu}(\tau)$.
 
 We want to construct a perturbative expansion of the {Maxwell} algebra by combining its contractions with the method of Lie algebra expansions~\cite{Hatsuda:2001pp,Boulanger:2002bt,deAzcarraga:2002xi,Izaurieta:2006zz,deAzcarraga:2007et,Bergshoeff:2019ctr}.
 The contraction of relevance in this paper can be obtained by a suitable rescaling of the generators. Starting from an algebra $\mathfrak{g}$ with generators $t_\alpha$ and structure constants $f_{\alpha\beta}{}^\gamma$, we define a new Lie algebra $\mf{g}_\omega$ for each $\omega> 0$ by rescaling some generators homogeneously with powers of $\omega$. For any $\omega> 0$ this is an invertible definition and the resulting Lie algebra $\mf{g}_\omega$ is isomorphic to the starting one.
The contracted algebra $\mf{g}_{\omega\to\infty}$ is well-defined if the limit $\omega\to \infty$ makes sense and the algebra is typically no longer isomorphic to $\mf{g}$ and one cannot necessarily invert the contraction process.
 
It is possible to keep track of information of the original algebra $\mf{g}$ by a perturbative scheme known as Lie algebra expansion.
To each generator $t_\alpha$ of $\mf{g}$ one associates a formal power series
 \begin{align}
 	\label{eq:tan}
 	t_\alpha \to \sum_{n\geq 0} t_\alpha \otimes \lambda^{n_0(\alpha) + n} = \sum_{n\geq 0} t_\alpha^{(n)} \,,
 \end{align}
 where the offset $n_0(\alpha)$ can depend on the generator.
 This produces an infinite-dimensional algebra out of $\mf{g}$ with generators $t_\alpha^{(n)}$. 
The label $n$ is to be thought of as the $n$th order perturbative expansion in the parameter $\lambda$ and the offset has to be chosen in such a way that the commutator of order $m$ with order $n$ only contains generators of order $\geq m{+}n$. The commutator is here defined by combining the commutator on $\mf{g}$ with the product of formal power series in $\lambda$.

At lowest order, the commutators involving of the $t_\alpha^{(0)}$ then can be arranged to be those of the contracted algebra, but the higher terms capture the perturbative expansion of the original algebra $\mathfrak{g}$. 

 A perturbative expansion for the {Poincaré} algebra adapted to the Galilean contraction was presented in \cite{Gomis:2019sqv}, obtaining an infinite-dimensional algebra $ \mathfrak{G}_\infty $. 
 
In Sections~\ref{sec:TDelec} and~\ref{sec:TDmag}, we will show the explicit contractions and constructions of the expanded algebra relevant to the electric and magnetic limits of electro-magnetism. In Appendix~\ref{app:free} we also show how this construction can be embedded in an even more general construction of infinite-dimensional \textit{free} Lie algebras in the magnetic and electric limits.

\subsection{Electric case}
\label{sec:TDelec}

The electric limit $ \mathfrak{E}$ of the Maxwell algebra is obtained by separating space and time indices and performing the following contraction of~\eqref{eq:Max}, see~\cite{Barducci:2019fjc,Gomis:2019fdh}:
\begin{subequations}
\begin{align} 
	\tilde{M}_{i j} & =M_{i j}, & \tilde{G}_i & =\frac{1}{\omega} M_{i 0} ,\\ \tilde{H} & =\omega  P_0, & \tilde{P}_i & = P_i ,\\ \tilde{Z}_{i j} & =\omega^2 Z_{i j}, & \tilde{Z}_i & =\omega Z_{0 i}.
\end{align}
\end{subequations}
The contracted algebra in the limit $\omega\to\infty$ has the commutation relations
\begin{subequations}
\begin{align} 
	{\left[\tilde{G}_i, \tilde{P}_j\right] } & =0, & {\left[\tilde{M}_{i j}, \tilde{P}_k\right] } & =2 \delta_{k[j} \tilde{P}_{i]}, & {\left[\tilde{G}_i, \tilde{Z}_j\right] } & =0, \\ {\left[\tilde{H}, \tilde{G}_i\right] } & =\tilde{P}_i, & {\left[\tilde{M}_{i j}, \tilde{G}_k\right] } & =2 \delta_{k[j} \tilde{G}_{i]}, &  \left[\tilde{P}_i, \tilde{P}_j\right]&=0,  \\ {\left[\tilde{H}, \tilde{P}_i\right] } & =\tilde{Z}_i, & {\left[\tilde{M}_{i j}, \tilde{Z}_k\right] } & =2 \delta_{k[j} \tilde{Z}_{i]}, & {\left[\tilde{G}_k, \tilde{Z}_{i j}\right] } & =2 \delta_{k[i} \tilde{Z}_{j]}, \\ 
	{\left[\tilde{G}_i, \tilde{G}_j\right] } & =0, & {\left[\tilde{M}_{i j}, \tilde{Z}_{k l}\right] } & =-4 \delta_{[i[l} \tilde{Z}_{k] j]}. & & 
\end{align}
\end{subequations}

We will construct the infinite-dimensional algebra $ \mathfrak{E}_{\infty} $ via the method of Lie algebra expansion by a semigroup~\cite{Hatsuda:2001pp,Boulanger:2002bt,deAzcarraga:2002xi,Izaurieta:2006zz,deAzcarraga:2007et,Bergshoeff:2019ctr}, $ S_E^{(\infty)} $ as follows.
Decompose the relativistic Maxwell algebra into~\cite{Gomis:2019nih}
\begin{align}
V_0=\{M_{ij},P_0,Z_{ij}\} \quad\text{and}\quad V_1=\{M_{0i},P_i,Z_{0i}\}\,,
\end{align}
which is a $ \mathbb{Z}_2$-grading:
\begin{align}
[V_0, V_0] \subset V_0 \,,\quad [V_0,V_1]\subset V_1\,, \quad [V_1,V_1]\subset V_0\,.
\end{align}
Using the resonant semigroups $ S_0^{(\infty)}= \{\lambda_{2m}|m=0,1,\dots\}  $ and $ S_1^{(\infty)}= \{\lambda_{2m+1}|m=0,1,\dots\}  $, we can construct the expanded algebra $V_0\otimes S_0^{(\infty)}\oplus  V_1\otimes S_1^{(\infty)}$ with generators 
\begin{subequations}
\begin{align}
	J^{(m)}_{ij}={M}_{ij}\otimes \lambda_{2m},&&H^{(m)}={P}_0\otimes \lambda_{2m} ,&& 	Z^{(m)}_{ij}={Z}_{ij}\otimes \lambda_{2m},\\
	G^{(m)}_{i}={M}_{0i}\otimes \lambda_{2m+1},&&P^{(m)}_i={P}_i\otimes \lambda_{2m+1}, && 	Z^{(m)}_{i}={Z}_{0i}\otimes \lambda_{2m+1}.
\end{align}
\end{subequations}
and commutation relations
\begin{subequations}
\label{eq:algelec}
\begin{align}
	\label{elec1}
	[J^{(m)}_{ij},J^{(n)}_{kl}]&=4\delta_{[i[k}J^{(m+n)}_{l]j]}, & \left[J^{(m)}_{i j}, P^{(n)}_k\right]&=-2 \delta_{k[i} P^{(m+n)}_{j]} , & 	[J^{(m)}_{ij},Z^{(n)}_{kl}]&=4\delta_{[i[k}Z^{(m+n)}_{l]j]},
	\\
	{\left[J^{(m)}_{i j}, G^{(n)}_k\right] }  &=-2 \delta_{k[i} G^{(m+n)}_{j]}, & \left[G^{(m)}_i, H^{(n)}\right]&=-P^{(m+n)}_i ,& 	[J^{(m)}_{ij},Z^{(n)}_{k}] &=2\delta_{k[j}Z^{(m+n)}_{i]},\\
	{\left[G^{(m)}_i, P^{(n)}_j\right] }  &=\delta_{ij}H^{(m+n+1)} , & \left[G^{(m)}_i, G^{(n)}_j\right]  &= J^{(m+n+1)}_{ij} ,& 	\left[Z^{(m)}_{i j}, G^{(n)}_k\right]   &=-2 \delta_{k[i} Z^{(m+n)}_{j]},\\
	\left[G^{(m)}_{i}, Z^{(n)}_j\right]   &=Z^{(m+n+1)}_{ij}, & \left[P^{(m)}_{i}, H^{(n)}\right]  &=-Z^{(m+n)}_{i}, & \left[P^{(m)}_{i}, P^{(n)}_j \right]   &=Z^{(m+n+1)}_{ij}.
	\label{elec2}
\end{align}
\end{subequations}
As noted in Appendix~\ref{app:free}, these expansions can be obtained as particular quotients of suitable Galilean free Lie algebras. Note that quotienting by the ideal generated by all generators of levels $ m\geq1 $ we recover the electric Maxwell algebra $\mathfrak{E}$.

\medskip

Our next aim is to construct a dynamical model that is invariant under the infinite-dimen\-sional $\mathfrak{E}_\infty$. This will be modelled after the charged particle~\eqref{eq:MaxL}.

To make the connection more transparent, we let $ \lambda_{2m}=c^{-2m} $ and $ \lambda_{2m+1}=c^{-2m-1}$ in the above expansion, and introducing convenient factors of $ c $. The generators of the electric Maxwell algebra then read
\begin{subequations}
	\label{eq:emg}
	\begin{align}
		H^{(m)} &= P_0\otimes c^{-2m+1}, & P_i^{(m)}&= P_i\otimes c^{-2m} ,& G_i^{(m)}&= M_{0i}\otimes c^{-2m-1},\\
		J_{ij}^{(m)}&= M_{ij}\otimes c^{-2m}, & Z_i^{(m)}&= Z_{0i}\otimes c^{-2m+1}, & Z^{(m)}_{ij}&= Z_{ij}\otimes c^{-2m+2}.
	\end{align}
\end{subequations}
If we give dimensions $L^{-1}$ to the relativistic translations, and no units to relativistic Lorentz transformations, then, using the above definitions, the level $ 0 $ generators, which we wish to identify with the Galilean limit of the Maxwell generators, have the following units $[H^{(0)}]=T^{-1}$ $[P_i^{(0)}]=L^{-1}$, $[G_i^{(0)}]=T L^{-1}$, $[J_{ij}^{(0)}]=1$, $[Z_i^{(0)}]=L^{-1}T^{-1}$ and $[Z_{ij}^{(0)}]=T^{-2}$.
The generators~\eqref{eq:emg} satisfy the commutation relations~\eqref{eq:algelec}.

{}From the algebra we define
 the homogeneous  space with generalised coordinates on which $ \mathfrak{E}_\infty $ acts, by quotienting by generalised ``Lorentz" generators: $ G_i^{(m)},J_{ij}^{(m)} $, i.e. the formal coset 
\begin{equation}
 \exp \mathfrak{E}_\infty / \exp \mathfrak{L}_\infty\,.
 \end{equation}
 Introduce in this space the coordinates $ x^i_{(m)},\,t_{(m)},\,\theta^i_{(m)},\,\phi^{ij}_{(m)} $ associated to $ P_i^{(m)}$, $H^{(m)}$, $Z_i^{(m)}$, $Z_{ij}^{(m)} $, respectively. In this way, $ x^i_{(0)} $ and $ t_{(0)} $ have units of length and time respectively, whereas $ \theta^i_{(0)} $ has units of $ LT $, and $ \phi^{ij}_{(0)} $ of $ T^{2} $. If we want $ \theta^i $ and $ \phi^{ij} $ to have interpretation as inertial momentum~\cite{Dixon:1967}, we need to add factors of $ c $ through the collective coordinates.

The infinitesimal action of a general element of the algebra of the form: 
\begin{align}
\sum_{n=0}^\infty \left( \epsilon_{(n)}H^{(n)}+\epsilon^i_{(n)}P_i^{(n)}+v^i_{(n)}G_i^{(n)}+\varepsilon^i_{(n)}Z_i^{(n)}+\varepsilon^{ij}_{(n)}Z_{ij}^{(n)}\right)
\end{align}
on the generalised space of coordinates is given by
\begin{subequations}
\label{eq:transelec}
\begin{align}
	\label{transele1}
	\delta t_{(m)}&=\epsilon_{(m)}+\sum_{n=0}^{m-1} v^i_{(m-n-1)}x^{(n)}_i\\
	\delta x^i_{(m)}&=\epsilon^i_{(m)}+\sum_{n=0}^m v^i_{(m-n)}t_{(n)} \\
	\delta \theta^i_{(m)}&=\varepsilon^i_{(m)}+\sum_{n=0}^m \frac 12 \epsilon^i_{(m-n)}t_{(n)}-\frac12 \epsilon_{(m-n)}x^i_{(n)}-2v^k_{(m-n)}\phi^{i}_{k(n)} \\
	\delta \phi^{ij}_{(m)}&=\varepsilon^{ij}_{(m)}+ \sum_{n=0}^{m-1} \epsilon^{[i}_{(m-n-1)}x^{j]}_{(n)}-2v^{[i}_{(m-n-1)}\theta^{j]}_{(n)}
	\label{transele2}
\end{align}
\end{subequations}

The connection to~\eqref{eq:MaxL} is via the collective coordinates
\begin{align}
	X^i&=\sum_{m=0}^\infty c^{-2m}x^i_{(m)},& T&=\sum_{m=0}^{\infty}c^{-2m}t_{(m)}, & \Theta^i &= \sum_{m=0}^\infty c^{-2m+1}\theta^i_{(m)}, &  \Phi^{ij} &= \sum_{m=0}^\infty c^{-2m+2}\phi^{ij}_{(m)},\nn\\
	&& F_{0i}&=\sum_{m=0}^\infty c^{-2m-1}f_{0i}^{(m)}, &F_{ij}&=\sum_{m=0}^\infty c^{-2m-2}f_{ij}^{(m)}. 
\end{align}
They give $ \Theta^i $ and $ \Phi^{ij} $ units of inertial momenta. With these definitions, $ f_{0i}^{(0)} $ already has dimensions of electric field, and $ f_{ij}^{(0)} $ dimensions of magnetic field.

Plugging  these collective coordinates into~\eqref{eq:MaxL} and grouping in powers of $ 1/c^2 $, we obtain the following expansion of the associated action $S= \sum_{m=0}^\infty S_{(m)}$ with $S_{(m)}$ proportional to $c^{2-2m}$:
{\allowdisplaybreaks
\begin{subequations}
\begin{align}
	S_{(0)}&=-mc^2\int d\tau \left[\dot{t}_{(0)}\right]\\
	S_{(1)}&=\int d\tau\left\{-m \left[\dot{t}_{(1)}-\frac{\dot{x}^2_{(0)}}{2\dot{t}_{(0)}}\right]-f_{0i}^{(0)}\left(\dot{\theta}^i_{(0)}+\frac12 \left(\dot{t}_{(0)}x^i_{(0)}-\dot{x}^i_{(0)}t_{(0)}\right)\right)-\frac 12 f_{ij}^{(0)}\left(\dot{\phi}^{ij}_{(0)}\right)\right\}\,, \\
	S_{(2)}&=\frac 1 {c^2}\int d\tau\left\{-m \left[\dot{t}_{(2)}-\frac{\dot{x}^i_{(0)}\dot{x}^j_{(1)}\delta_{ij}}{\dot{t}_{(0)}}+\frac{\dot{t}_{(1)}\dot{x}^2_{(0)}}{2\dot{t}^2_{(0)}}-	\frac{\dot{x}^4_{(0)}}{8\dot{t}^3_{(0)}}\right]\right.\nn\\*
	&\quad -f_{0i}^{(0)} \left(\dot{\theta}^i_{(1)}+\frac 12 \left(\dot{t}_{(1)}x^i_{(0)}+x^i_{(1)}\dot{t}_{(0)}-\dot{x}^i_{(0)}t_{(1)}-t_{(0)}\dot{x}^i_{(1)}\right)\right)\\*
	&\quad -f_{0i}^{(1)} \left(\dot{\theta}^i_{(0)}{+}\frac12 \left(\dot{t}_{(0)}x^i_{(0)}{-}\dot{x}^i_{(0)}t_{(0)}\right)\right)-\left.\frac 12 f_{ij}^{(0)}\left(\dot{\phi}^{ij}_{(1)}{+}\frac 12\left(\dot{x}^i_{(0)}x^j_{(0)}{-}\dot{x}^j_{(0)}x^i_{(0)}\right)\right){-} \frac 12 f_{ij}^{(1)}\dot{\phi}^{ij}_{(0)}\right\}\nn\\*
	&\dots \nonumber
\end{align}
\end{subequations}
These actions are invariant under the transformations~\eqref{eq:transelec}, if they are supplemented by the transformation  
\begin{align}
\delta f_{0i }^{(m)}  &=-\sum_{n=0}^{m-1}v^j_{(n)}f^{(m-n-1)}_{ij}
\,,\nn\\
\delta f_{ij }^{(m)}  &=2 \sum_{n=0}^m v^{(n)}_{[i}f^{(m-n)}_{0|j]}\,.
\end{align}
}

The action $S_{(0)}$ is a total derivative with trivial dynamics, the action $S_{(1)}$ gives the strict electric limit of the charged particle. Since we are interested in corrections, we investigate the equations of motion of $ S_{(2)} $ that read:
\begin{subequations}
\label{eq:xxx}
	\begin{align}
		\frac{m}2 \frac{d}{d\tau}\left( \frac{\dot{x}^2_{(0)}}{\dot{t}^2_{(0)}}\right)&=-f_{0i}^{(0)}\dot{x}^i_{(0)}\,,\\
		m\frac{d}{d\tau} \left(\frac{\dot{x}_{(0)}}{\dot{t}_{(0)}}\right)&=- f_{0i}^{(0)}\dot{t}_{(0)}\,,\\
		m\frac d {d\tau }\left(\frac{\dot{\vec{x}}_{(1)}\cdot\dot{\vec{x}}_{(0)}}{\dot{t}^2_{(0)}}-\frac{\dot{t}_{(1)}\dot{x}_{(0)}^2}{\dot{t}^3_{(0)}}+\frac 38 \frac{\dot{x}^4_{(0)}}{\dot{t}^4_{(0)}}\right)&=-f_{0i}^{(1)}\dot{x}^i_{(0)}-f_{0i}^{(0)}\dot{x}^i_{(1)}\,,\\
		m\frac d {d\tau }\left(\frac{\dot{x}_{i(1)}}{\dot{t}_{(0)}}-\frac{\dot{t}_{(1)}\dot{x}_{i(0)}}{\dot{t}^2_{(0)}}+\frac 12 \frac{\dot{\vec{x}}^2_{(0)}\dot{x}_{i(0)}}{\dot{t}^3_{(0)}}\right)&=-f_{0i}^{(1)}\dot{t}_{(0)}-f_{0i}^{(0)}\dot{t}_{(1)}+ f_{ij}^{(0)}\dot{x}^j_{(0)}\,.
	\end{align}
\end{subequations}
In deriving these equations, we have used the equations obtained by varying the fields $\theta^i_{(m)}$ and $\phi^{ij}_{(m)}$ that read
\begin{align}
\dot{f}_{ij}^{(m)} = \dot{f}_{0i}^{(m)} = 0
\end{align}
and force the components $f_{ij}^{(m)}$ and $f_{0i}^{(m)}$ to be constants.

For these $\tau$-independent quantities we see that~\eqref{eq:xxx} agree with the equations~\eqref{eq:LFelec} under the identification $ \tilde{F}^{ti}_{(m)}=f^{0i}_{(m)}\, , F^{ij}_{(m+1)}=f^{ij}_{(m)} $ and after raising/lowering indices.

\subsection{Magnetic case}
\label{sec:TDmag}

The magnetic limit of the Maxwell algebra is instead obtained from rescaling the generators as follows~\cite{Barducci:2019fjc}
\begin{subequations}
\begin{align} 
	\tilde{M}_{i j} & =M_{i j}, & \tilde{G}_i & =\frac{1}{\omega} M_{i 0} ,\\
	 \tilde{H} & = \omega P_0, & \tilde{P}_i & = P_i ,\\ 
	 \tilde{Z}_{i j} & = Z_{i j}, & \tilde{Z}_i & =\omega Z_{0 i}.
\end{align}
\end{subequations}
The new commutation relations, after taking the limit $ \omega\rightarrow \infty $ read
\begin{subequations}
\begin{align}
	 \left[\tilde{G}_i, \tilde{P}_j\right]&=0, & \left[\tilde{M}_{i j}, \tilde{P}_k\right]&=2 \delta_{k[j} \tilde{P}_{i]}, &\left[\tilde{G}_i, \tilde{Z}_j\right]&=-\tilde{Z}_{i j}, \\
	\left[\tilde{H}, \tilde{G}_i\right] &=\tilde{P}_i, & \left[\tilde{M}_{i j}, \tilde{G}_k\right]&=2 \delta_{k[j} \tilde{G}_{i]}, &\left[\tilde{P}_i, \tilde{P}_j\right]&=\tilde{Z}_{i j}, \\
	\left[\tilde{H}, \tilde{P}_i\right] &=\tilde{Z}_i, &\left[\tilde{M}_{i j}, \tilde{Z}_k\right]&=2 \delta_{k[j} \tilde{Z}_{i]}, &\left[\tilde{G}_k, \tilde{Z}_{i j}\right]&=0, \\
	\left[\tilde{G}_i, \tilde{G}_j\right]&=0, &\left[\tilde{M}_{i j}, \tilde{Z}_{k l}\right]&=-4 \delta_{[i[l} \tilde{Z}_{k] j]} .
\end{align}
\end{subequations}

The infinite-dimensional algebra $ \mathfrak{M}_\infty $ will be constructed as an expansion of the Maxwell algebra by a semigroup $ S_E^{(\infty)} $ as follows. We decompose the generators of the Maxwell algebra into~\cite{Izaurieta:2006zz}
\begin{align}
V_0=\{M_{ij},H\},\quad V_1=\{G_i,P_i,Z_i\},\quad V_2=\{Z_{ij}\},
\end{align}
with
\begin{align}
[V_0, V_i] \subset V_i\,, \quad [V_1,V_1]\subset V_0\oplus V_2\,,\quad [V_1,V_2]\subset V_1\,,\quad [V_2, V_2]\subset V_0\,.
\end{align}
A resonant semigroup will be, $ S_{0}^{(\infty)}=\{\lambda_{2m}|m=0,1,\dots \} $, $ S_{1}^{(\infty)}=\{\lambda_{2m+1}|m=0,1,\dots \} $ and $ S_{2}^{(\infty)}=\{\lambda_{2m+2}|m=0,1,\dots \} $.	
Defining the new expanded algebra:
\begin{align}
V_0\otimes S_0^{(\infty)} \oplus V_1\otimes S_1^{(\infty)} \oplus V_2\otimes S_2^{(\infty)},
\end{align}
whose generators are
\begin{subequations}
\begin{align}
	J^{(m)}_{ij}&={M}_{ij}\otimes \lambda_{2m},&H^{(m)}&={P}_0\otimes \lambda_{2m}, & 	Z^{(m)}_{ij}&={Z}_{ij}\otimes \lambda_{2m+2},\\
	G^{(m)}_{i}&={M}_{0i}\otimes \lambda_{2m+1},&P^{(m)}_i&={P}_i\otimes \lambda_{2m+1}, & 	Z^{(m)}_{i}&={Z}_{0i}\otimes \lambda_{2m+1}.
\end{align}
\end{subequations}
The expanded algebra will be denoted by $ \mathfrak{M}_\infty $ and has the following commutation relations:
\begin{subequations}
\label{eq:algmag}
\begin{align} \label{eq:Mag1}
	[J^{(m)}_{ij},J^{(n)}_{kl}]&=4\delta_{[i[k}J^{(m+n)}_{l]j]}, & \left[J^{(m)}_{i j}, P^{(n)}_k\right]&=-2 \delta_{k[i} P^{(m+n)}_{j]} , & 	[J^{(m)}_{ij},Z^{(n)}_{kl}]&=4\delta_{[i[k}Z^{(m+n)}_{l]j]},\\
	\left[J^{(m)}_{i j}, G^{(n)}_k\right]   &=-2 \delta_{k[i} G^{(m+n)}_{j]}, & \left[H^{(m)}, G^{(n)}_i	\right]&=P^{(m+n)}_i , & 	[J^{(m)}_{ij},Z^{(n)}_{k}]&=2\delta_{k[j}Z^{(m+n)}_{i]}\\
	\left[G^{(m)}_i, P^{(n)}_j\right] & =\delta_{ij}H^{(m+n+1)},  & \left[G^{(m)}_i, G^{(n)}_j\right] & = J^{(m+n+1)}_{ij}, & 	\left[G^{(m)}_{k}, Z^{(n)}_{ij}\right] &  =2 \delta_{k[i} Z^{(m+n+1)}_{j]},\\
	{\left[G^{(m)}_{i}, Z^{(n)}_j\right] }  &=Z^{(m+n)}_{ij}, & {\left[H^{(m)}, P^{(n)}_i\right] }  &=Z^{(m+n)}_{i}, & \left[P^{(m)}_{i}, P^{(n)}_j \right]   &=Z^{(m+n)}_{ij},\\\label{eq:Mag2}
	\left[G^{(m)}_i, P^{(n)}_j\right] & =\delta_{ij}H^{(m+n+1)},  &\left[H^{(m)}, G^{(n)}_i	\right]&=P^{(m+n)}_i, & 	\left[G^{(m)}_{k}, Z^{(n)}_{ij}\right] &  =2 \delta_{k[i} Z^{(m+n+1)}_{j]},\\
	{\left[G^{(m)}_{i}, Z^{(n)}_j\right] }  &=Z^{(m+n)}_{ij}, & {\left[H^{(m)}, P^{(n)}_i\right] }  &=Z^{(m+n)}_{i}, & \left[P^{(m)}_{i}, P^{(n)}_j \right]   &=Z^{(m+n)}_{ij},
\end{align}
\end{subequations}
One recovers the strict limit of the magnetic Galilei algebra when setting all generators from level $ m\geq1 $ to zero.

\medskip

A dynamical model with this symmetry can be found in the same as for the electric limit. We first set $\lambda_{2m}=c^{-2m}$ and $\lambda_{2m+1}=c^{-2m-1}$ and relabel the generators of the infinite-dimensional magnetic Maxwell algebra $\mathfrak{M}_\infty$:
\begin{subequations}
	\begin{align}
		H^{(m)} &= P_0\otimes c^{-2m+1}, & P_i^{(m)}&= P_i\otimes c^{-2m} ,& G_i^{(m)}&= M_{0i}\otimes c^{-2m-1},\\
		J_{ij}^{(m)}&= M_{ij}\otimes c^{-2m}, & Z_i^{(m)}&= Z_{0i}\otimes c^{-2m+1}, & Z^{(m)}_{ij}&= Z_{ij}\otimes c^{-2m}.
	\end{align}
\end{subequations}
They satisfy the relations~\eqref{eq:algmag}.

In this the generalised  homogeneous space on which $ \mathfrak{M}_\infty $ acts non-linearly is given as the formal coset 
\begin{align}
 \exp \mathfrak{M}_\infty / \exp \mathfrak{L}_\infty,
 \end{align}
by quotienting by the generalised Lorentz generators. We introduce in this space the coordinates $ x^i_{(m)},\,t_{(m)},\,\theta^i_{(m)},\,\phi^{ij}_{(m)} $, associated to $ P_i^{(m)},H^{(m)},Z_i^{(m)},Z_{ij}^{(m)} $, respectively.

The infinitesimal action of a general element of the $ \mathfrak{M}_\infty $ algebra of the form: 
\begin{align}
\sum_{n=0}^\infty\left( \epsilon_{(n)}H^{(n)}+\epsilon^i_{(n)}P_i^{(n)}+v^i_{(n)}G_i^{(n)}+\varepsilon^i_{(n)}Z_i^{(n)}+\varepsilon^{ij}_{(n)}Z_{ij}^{(n)}\right)
\end{align}
on the generalised space of coordinates is given by
\begin{subequations}
\label{eq:transmag}
\begin{align}
	\label{transmag1}
	\delta t_{(m)}&=\epsilon_{(m)}+\sum_{n=0}^{m-1} v^i_{(m-n-1)}x^{(n)}_i,\\
	\delta x^i_{(m)}&=\epsilon^i_{(m)}+\sum_{n=0}^mv^i_{(m-n)}t_{(n)} ,\\
	\delta \theta^i_{(m)}&=\varepsilon^i_{(m)}+\sum_{n=0}^{m} \frac 12 \epsilon^i_{(m-n)}t_{(n)}-\frac12 \epsilon_{(m-n)}x^i_{(n)}-2v^k_{(m-n-1)}\phi^{i}_{k(n)}, \\
	\delta \phi^{ij}_{(m)}&=\varepsilon^{ij}_{(m)}+ \sum_{n=0}^m \epsilon^{[i}_{(m-n)}x^{j]}_{(n)}-2v^{[i}_{(m-n)}\theta^{j]}_{(n)}
	\label{transmag2}.
\end{align}
\end{subequations}

With the collective coordinates
\begin{align}
	X^i&=\sum_{m=0}^\infty c^{-2m}x^i_{(m)}, & T&=\sum_{m=0}^{\infty}c^{-2m}t_{(m)} ,& \Theta^i &= \sum_{m=0}^\infty c^{-2m+1}\theta^i_{(m)}, &  \Phi^{ij} &= \sum_{m=0}^\infty c^{-2m}\phi^{ij}_{(m)},\nn\\
	&& F_{0i}&=\sum_{m=0}^\infty c^{-2m-1}f_{0i}^{(m)},& F_{ij}&=\sum_{m=0}^\infty c^{-2m}f_{ij}^{(m)}.& 
\end{align}
one can expand~\eqref{eq:MaxL} and obtain the first few expanded actions as
\begin{subequations}
\begin{align}
	S_{(0)} &=-mc^2\int d\tau \left[\dot{t}_{(0)}\right],\\
	S_{(1)} &=\int d\tau\left\{ -m\left[\dot{t}_{(1)}{-}\frac{\dot{x}^2_{(0)}}{2\dot{t}_{(0)}}\right]-f_{0i}^{(0)}\left(\dot{\theta}^i_{(0)}{+}\frac12 \left(\dot{t}_{(0)}x^i_{(0)}{-}\dot{x}^i_{(0)}t_{(0)}\right)\right){-}\frac 12 f_{ij}^{(0)}\left(\dot{\phi}^{ij}_{(0)}{+} \dot{x}^{[i}_{(0)}x^{j]}_{(0)}\right)\right\},\\
	S_{(2)}& =\frac 1 {c^2}\int d\tau\left\{m \left[-\dot{t}_{(2)}{+}\frac{\dot{x}^i_{(0)}\dot{x}^j_{(1)}\delta_{ij}}{\dot{t}_{(0)}}{-}\frac{\dot{t}_{(1)}\dot{x}^2_{(0)}}{2\dot{t}^2_{(0)}}{+}\frac{\dot{x}^4_{(0)}}{8\dot{t}^3_{(0)}}\right]-f_{0i}^{(1)}\left(\dot{\theta}^i_{(0)}{+}\frac12 \left(\dot{t}_{(0)}x^i_{(0)}{-}\dot{x}^i_{(0)}t_{(0)}\right)\right)\right.\nn\\
	&\quad-\left.f_{0i}^{(0)} \left(\dot{\theta}^i_{(1)}+\frac 12 \left(\dot{t}_{(1)}x^i_{(0)}+x^i_{(1)}\dot{t}_{(0)}-\dot{x}^i_{(0)}t_{(1)}-t_{(0)}\dot{x}^i_{(1)}\right)\right)\right.\nn\\
	&\quad -f_{ij}^{(0)}\left(\dot{\phi}^{ij}_{(1)}+\left(\dot{x}^{[i}_{(1)}x^{j]}_{(0)}+\dot{x}^{[j}_{(0)}x^{i]}_{(1)}\right)\right)-f_{ij}^{(1)}\left(\dot{\phi}^{ij}_{(0)}+\dot{x}^{[i}_{(0)}x^{j]}_{(0)}\right)\Big\}\\
	&\dots\nn
\end{align}
\end{subequations}
Each term in the expansion is invariant under~\eqref{eq:transmag}, if supplemented by
\begin{align}
	\delta f_{0i }^{(m)}  &=-\sum_{n=0}^{m-1}v^j_{(n)}f^{(m-n)}_{ij}
	\,,\nn\\
	\delta f_{ij }^{(m)}  &=2 \sum_{n=0}^m v^{(n)}_{[i}f^{(m-n-1)}_{0|j]}\,.
\end{align}
Notice that the transformations in the magnetic limit differ in the summation range from the ones in the electric limit.

The action $S_{(0)}$ is a total derivative and $S_{(1)}$ is the strict magnetic limit of a particle. 
The equations of motion for $ S_{(2)} $ read:
\begin{subequations}
\begin{align}
	m\frac 12 \frac{d}{d\tau }\left( \frac{\dot{x}^2_{(0)}}{\dot{t}^2_{(0)}}\right)&=-f_{0i}^{(0)}\dot{x}^i_{(0)},\\
	m\frac d {d\tau}\left( \frac{\dot{x}_i^{(0)}}{\dot{t}_{(0)}}\right)&=-f_{0i}^{(0)}\dot{t}_{(0)}+ f_{ij}^{(0)}\dot{x}^j_{(0)},\\
	m\frac{d}{d\tau}\left(\frac{\dot{\vec{x}}_{(0)}\cdot\dot{\vec{x}}_{(1)}}{\dot{t}^2_{(0)}}-\frac{\dot{t}_{(1)}\dot{x}^2_{(0)}}{\dot{t}_{(0)}^3}+\frac 3 8 \frac{\dot{x}^4_{(0)}}{\dot{t}_{(0)}^4} \right)&= -f_{0i}^{(1)}\dot{x}^i_{(0)}-f_{0i}^{(0)}\dot{x}^i_{(1)},\\
	m\frac d {d\tau}\left(\frac{\dot{x}_{i(1)}}{\dot{t}_{(0)}}-\frac{\dot{x}_{i(0)}\dot{t}_{(1)}}{\dot{t}^2_{(0)}}+\frac{\dot{x}^2_{(0)}\dot{x}_{i(0)}}{2\dot{t}^3_{(0)}}\right)&=-  f_{0i}^{(1)}\dot{t}_{(0)}-f_{0i}^{(0)}\dot{t}_{(1)}+ f_{ij}^{{(0)}}\dot{x}^j_{(1)}+ f_{ij}^{{(1)}}\dot{x}^j_{(0)}.
\end{align}
\end{subequations}
They again coincide with~\eqref{eq:expLF} after identifying $ \tilde{F}^{ti}_{(m)}=f^{0i}_{(m)} \,, F^{ij}_{(m)}=f^{ij}_{(m)}$, which shows agreement between the bottom-up and top-down approach.

The role of the $ \theta^i$ and $ \phi^{ij} $ variables is twofold. First, they appear in the Lagrangian as Lagrange multipliers to ensure that the new variables $ f_{0i}(\tau) $ and $ f_{ij}(\tau) $ remain constant on-shell. Second, they maintain invariance, as the fields $ f_{0i} $ and $ f_{ij} $ are the conjugate momenta of $ \theta^i $ and $ \phi^{ij} $, and therefore their transformations laws are determined by the latter.

The boost transformations again act indecomposably on the expanded coordinates and fields as well as on the equations of motion.

\section{Carroll limit}
\label{sec:Carroll}

We can repeat the same analysis for the Carroll limit of a charged particle. In order to have any non-trivial dynamics we consider a Carroll tachyon~\cite{deBoer:2021jej,Gomis:2022spp} since an ordinary Carroll particle cannot move. The steps are very similar to above and we shall be brief. 

\subsection{Bottom-up approach}

The starting point is the relativistic Lorentz equation~\eqref{Lorentz}, but now for a tachyon we use Carroll time $s=C x^0$ in terms of the Carroll speed of light $C$ first introduced in~\cite{LevyLeblond:1965,Duval:2014uoa}.  Note that $s$ has dimensions $L^2/T$. Moreover, we have used the rescaled Carroll mass $\tilde{M}$ that satisfies $MC=mc=\tilde{M}$~\cite{Gomis:2022spp}. The separated relativistic equations are
\begin{subequations}
	\label{eq:MaxC}
	\begin{align}
		\tilde{M}\frac d {d\tau}\left(\frac{\dot{s}}{\sqrt{\dot{\vec{x}}^2}\sqrt{1-\frac{\dot{s}^2}{\dot{\vec{x}}^2C^2}}}\right)&=C^2\tilde{F}^{si}\dot{x}_i\\
		\tilde{M}\frac d {d\tau}\left(\frac{\dot{x}^i}{\sqrt{\dot{\vec{x}}^2}\sqrt{1-\frac{\dot{s}^2}{\dot{\vec{x}}^2C^2}}}\right)&={\tilde{F}^{si}}\dot{s}+{F^{ij}}\dot{x}_j\,,
	\end{align}
\end{subequations}
where we have introduced $  \tilde{F}^{si}=F^{0i}/C$. This definition is hinted at by analogy with the non-relativistic case, where the electric field is defined as $ \tilde{F}^{ti}=cF^{0i} $. In the Carroll case we define the electro-magnetic fields as 
\begin{equation}
	\label{def:Cfields}
	\tilde{F}^{si}=F^{0i}/C=\frac{\tilde{F}^{ti}}{cC} ,\; \tilde{F}^{ij}=F^{ij}
\end{equation}
These definitions are also suggested by the Carrollian electric contraction of the Maxwell algebra that we discuss below in~\eqref{eq:CarrollElecContraction}, where $ C^{-1} $ plays the role of the contraction parameter.
Even though this definition differs from the one used by~\cite{Duval:2014uoa}, it is completely equivalent, as they also satisfy the Carroll version of the Maxwell equations.

The fact that the spatial velocity appears with a positive sign under the square root despite our signature $(-+\ldots +)$ is due to the tachyonic nature of the particle. The equations~\eqref{eq:MaxC} possess full relativistic invariance, although not manifestly so.

In the limit $C\to \infty$, the analogue of the gamma factor expands as
\begin{align}
	\frac{1}{\sqrt{\dot{\vec{x}}^{\, 2} - \dot{s}^2 / C^2 }} = 
	\frac{1}{ \sqrt{\dot{\vec{x}}^{\, 2} }} + \frac{\dot{s}^2}{2 C^2  \dot{\vec{x}}^{\, 2} \sqrt{\dot{\vec{x}}^{\, 2} }} + \frac{3 \dot{s}^4}{8 C^4  (\dot{\vec{x}}^{\, 2})^2 \sqrt{\dot{\vec{x}}^{\, 2} }}  + \ldots 
\end{align}
We also consider an expansion of the coordinates and of the electro-magnetic field according to
\begin{subequations}
	\begin{align}
		\label{eq:coordC1}
		s &= \sum_{m=0}^\infty s_{(m)} C^{-2m} \,, & x^i &= \sum_{m=0}^\infty x^i_{(m)} C^{-2m}\,,\\
		\tilde{F}^{si} &= \sum_{m=0}^\infty \tilde{F}^{si}_{(m)} C^{-2m}\,, & 
		{F}^{ij} &= \sum_{m=0}^\infty {F}^{ij}_{(m)} C^{-2m} \,.
	\end{align}
\end{subequations}
Substituting this into~\eqref{eq:MaxC} we obtain in the lowest orders in $1/C^2$:
\begin{subequations}
	\label{eq:covC2}
	\begin{align}
	\label{cc1a}
		\tilde{F}^{si}_{(0)}\dot{x}_{(0)i}&=0,\,\\
	\label{cc1b}		
		\tilde{M} \frac{d}{d\tau} \left( \frac{\dot{x}^i_{(0)}}{\sqrt{\dot{\vec{x}}_{(0)}^2}}\right) &=\tilde{F}^{si}_{(0)}\dot{s}_{(0)}+{F}^{ij}_{(0)}\dot{x}_{(0)j}\,,\\
	\label{cc1c}
		\tilde{M} \frac{d}{d\tau} \left( \frac{\dot{s}_{(0)}}{\sqrt{\dot{\vec{x}}_{(0)}^2}}\right) &= \tilde{F}^{si}_{(0)} \dot{x}_{(1) i}+ \tilde{F}^{si}_{(1)} \dot{x}_{(0) i}\,,\\
	\label{cc1d}
		\tilde{M} \frac{d}{d\tau} \left( \frac{\dot{x}^i_{(1)}}{\sqrt{\dot{\vec{x}}_{(0)}^2}} - \frac{\dot{x}_{(0)}^i \dot{\vec{x}}_{(0)}\cdot \dot{\vec{x}}_{(1)}}{\dot{\vec{x}}_{(0)}^2 \sqrt{\dot{\vec{x}}_{(0)}^2} } + \frac{\dot{x}^i_{(0)} \dot{s}_{(0)}^2}{2\dot{\vec{x}}_{(0)}^2 \sqrt{\dot{\vec{x}}_{(0)}^2} }
		\right) &= \tilde{F}^{si}_{(1)} \dot{s}_{(0)} +\tilde{F}^{si}_{(0)} \dot{s}_{(1)} + {F}^{ij}_{(1)} \dot{x}_{(0) j} + {F}^{ij}_{(0)} \dot{x}_{(1) j} \,,\\
	\label{cc1e}
		M \frac{d}{d\tau} \left( \frac{\dot{s}_{(1)}}{\sqrt{\dot{\vec{x}}_{(0)}^2}}   - \frac{\dot{s}_{(0)} \dot{\vec{x}}_{(0)}\cdot \dot{\vec{x}}_{(1)}}{\dot{\vec{x}}_{(0)}^2 \sqrt{\dot{\vec{x}}_{(0)}^2}}  + \frac{\dot{s}_{(0)}^3}{2\dot{\vec{x}}_{(0)}^2 \sqrt{\dot{\vec{x}}_{(0)}^2} }\right) &= \tilde{F}^{si}_{(0)} \dot{x}_{(2) i}+\tilde{F}^{si}_{(1)} \dot{x}_{(1) i} +\tilde{F}^{si}_{(2)} \dot{x}_{(0) i}  \,.
	\end{align}
\end{subequations}
Note how as in the Galilei case, the different level equations are related via boosts as
\begin{equation}
\dots\xrightarrow{\beta_{(2)}^i}\eqref{cc1e}\xrightarrow{\beta_{(1)}^i}\eqref{cc1d}\xrightarrow{\beta_{(1)}^i}\eqref{cc1c}\xrightarrow{\beta_{(0)}^i}\eqref{cc1b}\xrightarrow{\beta_{(0)}^i}\eqref{cc1a}\xrightarrow{\beta_{(0)}^i}0\,.
\end{equation}
This also justifies the presence of the first equation~\eqref{cc1a}, arising at order $C^2$
 in the limit of~\eqref{eq:MaxC},  that might be surprising at first sight since it is not a second-order differential equation but a transversality constraint. 

We consider  the electric and magnetic limits of the electro-magnetic field.\footnote{One could also consider a pulse limit. This would require a $1/C$ expansion that proceeds very similarly to the Galilei case in Appendix~\ref{app:1c} and that we do not spell out.} Translating them in terms of the Carroll electric and magnetic fields~\eqref{def:Cfields}: 
\begin{align}
 \textbf{E}_{\text{Galilei}}\gg c\textbf{B}_{\text{Galilei}} \iff \textbf{B}_{\text{Carroll}}\ll C\textbf{E}_{\text{Carroll}}.
 \end{align} 
The magnetic field was naturally greater in the non-relativistic limit, whereas the electric field is the bigger one in the Carroll limit, hinting at a duality between the Carroll and Galilei limits. We will come back to this duality in Section~\ref{sec:duality}.

\subsubsection{Electric Carroll limit}

The electric limit can be implemented in our expansion by keeping $ \tilde{F}^{si}_{(0)}\neq 0 $ since $ \tilde{F}^{si}_{(0)}\gg F^{ij}_{(0)}/C $, whereas the magnetic limit will correspond to $ F^{ij}_{(0)}\neq 0 $ and $ \tilde{F}^{si}_{(0)}=0 $.

The equations of motion in the electric limit are formally the same as~\eqref{eq:covC2}. The first equation says that at the first level, the motion of the tachyon is perpendicular to the electric field. Note that this restriction is similar to a non-relativistic particle which moves perpendicular to the magnetic field.

They are invariant under the following transformations of the coordinates as well as the fields: 
\begin{align}
\label{eq:Cetrm}
	\delta s_{(n)} &= \epsilon_{(n)}+\sum_{m=0}^n \vec{\beta}_{(m)} \cdot \vec{x}_{(n-m)}\,,&
	\delta x_{(n)}^i &= \epsilon^i_{(n)}+\sum_{m=0}^{n-1} \beta_{(m)}^i s_{(n-m-1)}\,,\nn\\
	\delta \tilde{F}^{si}_{(n)} &= \sum_{m=0}^{n-1} \beta_{(m) j} {F}^{ji}_{(n-m-1)} 
	\,,&
	\delta {F}^{ij}_{(n)} &= -2 \sum_{m=0}^{n} \tilde{F}^{s[i}_{(m)} \beta^{j]}_{(n-m)} \,,
\end{align}
where we have defined the Carroll boost $ \beta^i=Cv^i/c $ and have performed an expansion $ \beta^i=\sum_{m=0}^\infty \beta^i_{(m)} C^{-2m} $.

\subsubsection{Magnetic Carroll limit}

The equations in the magnetic limit read:
{\allowdisplaybreaks
\begin{subequations}
	\label{eq:covC2mag}
	\begin{align}
		\tilde{M} \frac{d}{d\tau} \left( \frac{\dot{s}_{(0)}}{\sqrt{\dot{\vec{x}}_{(0)}^2}}\right) &= \tilde{F}^{si}_{(1)} \dot{x}_{(0) i}\,,\\	
			\tilde{M} \frac{d}{d\tau} \left( \frac{\dot{x}^i_{(0)}}{\sqrt{\dot{\vec{x}}_{(0)}^2}}\right) &={F}^{ij}_{(0)}\dot{x}_{(0)j}\,,\\
		M \frac{d}{d\tau} \left( \frac{\dot{s}_{(1)}}{\sqrt{\dot{\vec{x}}_{(0)}^2}}   - \frac{\dot{s}_{(0)} \dot{\vec{x}}_{(0)}\cdot \dot{\vec{x}}_{(1)}}{\dot{\vec{x}}_{(0)}^2 \sqrt{\dot{\vec{x}}_{(0)}^2}}  + \frac{\dot{s}_{(0)}^3}{2\dot{\vec{x}}_{(0)}^2 \sqrt{\dot{\vec{x}}_{(0)}^2} }\right) &= \tilde{F}^{si}_{(1)} \dot{x}_{(1) i} +\tilde{F}^{si}_{(2)} \dot{x}_{(0) i}  \,,\\
		\tilde{M} \frac{d}{d\tau} \left( \frac{\dot{x}^i_{(1)}}{\sqrt{\dot{\vec{x}}_{(0)}^2}} - \frac{\dot{x}_{(0)}^i \dot{\vec{x}}_{(0)}\cdot \dot{\vec{x}}_{(1)}}{\dot{\vec{x}}_{(0)}^2 \sqrt{\dot{\vec{x}}_{(0)}^2} } + \frac{\dot{x}^i_{(0)} \dot{s}_{(0)}^2}{2\dot{\vec{x}}_{(0)}^2 \sqrt{\dot{\vec{x}}_{(0)}^2} }
		\right) &= \tilde{F}^{si}_{(1)} \dot{s}_{(0)} + {F}^{ij}_{(1)} \dot{x}_{(0) j} + {F}^{ij}_{(0)} \dot{x}_{(1) j} \,.
	\end{align}
\end{subequations}
}

These are instead invariant under
\begin{align}
	\delta s_{(n)} &= \epsilon_{(n)}+\sum_{m=0}^n \vec{\beta}_{(m)} \cdot \vec{x}_{(n-m)}\,,&
	\delta x_{(n)}^i &= \epsilon^i_{(n)}+\sum_{m=0}^{n-1} \beta_{(m)}^i s_{(n-m-1)}\,,\nn\\
	\delta \tilde{F}^{si}_{(n)} &= \sum_{m=0}^{n} \beta_{(m) j} {F}^{ji}_{(n-m)} 
	\,,&
	\delta {F}^{ij}_{(n)} &= -2 \sum_{m=0}^{n-1} \tilde{F}^{s[i}_{(m)} \beta^{j]}_{(n-m-1)} \,.
\end{align}
They differ from~\eqref{eq:Cetrm} in summation ranges.

\subsection{Lie algebraic point of view}

In this section, we present the details of the expansion of Maxwell algebra in the different Carrollian limits. We also study the equations of motion in the top-down approach.

\subsubsection{Electric Carroll Maxwell}

The electric limit can be obtained via a contraction of the Maxwell algebra~\eqref{eq:Max}, see~\cite{Barducci:2019fjc}:
\begin{align} 
	\label{eq:CarrollElecContraction}
	\tilde{M}_{i j} & =M_{i j}, & \tilde{G}_i & =\frac{1}{\omega} M_{i 0} \,,\nn\\
	\tilde{H} & = \frac 1 \omega P_0, & \tilde{P}_i & = P_i \,,\nn\\ 
	\tilde{Z}_{i j} & = Z_{i j}, & \tilde{Z}_i & =\frac 1 \omega Z_{0 i}
\end{align}
and taking the limit $ \omega\rightarrow \infty $, the  contracted commutation relations become
\begin{subequations}
\begin{align} 
	{\left[\tilde{G}_i, \tilde{P}_j\right] } & =\delta_{ij}\tilde{H}, & {\left[\tilde{M}_{i j}, \tilde{P}_k\right] } & =2 \delta_{k[j} \tilde{P}_{i]}, & {\left[\tilde{G}_i, \tilde{Z}_j\right] } & =0, \\ 
	{\left[\tilde{H}, \tilde{G}_i\right] } & =0, & {\left[\tilde{M}_{i j}, \tilde{G}_k\right] } & =2 \delta_{k[j} \tilde{G}_{i]}, &  \left[\tilde{P}_i, \tilde{P}_j\right]&=\tilde{Z}_{ij},  \\
	{\left[\tilde{H}, \tilde{P}_i\right] } & =\tilde{Z}_{i}, & {\left[\tilde{M}_{i j}, \tilde{Z}_k\right] } & =2 \delta_{k[j} \tilde{Z}_{i]}, & {\left[\tilde{G}_k, \tilde{Z}_{i j}\right] } & =2\delta_{k[i}\tilde{Z}_{j]}, \\
	{\left[\tilde{G}_i, \tilde{G}_j\right] } & =0, & {\left[\tilde{M}_{i j}, \tilde{Z}_{k l}\right] } & =-4 \delta_{[i[l} \tilde{Z}_{k] j]} . & & 
\end{align}
\end{subequations}
For applying the expansion method to the electric Carroll we use the following division into subspaces~\cite{Gomis:2019nih} 
\begin{align}
	V_0=\{J_{ij},P_i,Z_{ij}\} \,,&& V_1=\{G_i,H,Z_i\}\,.
\end{align}
Then we define the generators
\begin{align}
	\label{def:ecGen}
	J^{(m)}_{ij}&=M_{ij}\otimes C^{-2m},& H^{(m)}&=P_0\otimes C^{-2m-1}, & 	Z^{(m)}_{ij}&=Z_{ij}\otimes C^{-2m},\nn\\
	G^{(m)}_{i}&=M_{0i}\otimes C^{-2m-1},&P^{(m)}_i&=P_i\otimes C^{-2m}, & Z^{(m)}_i&=Z_{0i}\otimes C^{-2m-1}.
\end{align}

The commutation relations of the infinite-dimensional algebra are	
\begin{subequations}
	\begin{align}
		\left[G_{i}^{(m)}, H^{(n)}\right]  & = P_{i}^{(m+n+1)}\,, &		\left[G_{i}^{(m)}, P_{j}^{(n)}\right]  &=\delta_{ij} H^{(m+n)} \,, \\
		\left[J_{ij}^{(m)}, P_k^{(n)}\right]  & =2 \delta_{k[j} P_{i]}^{(m+n)}\,, &
		\left[G_{i}^{(m)}, G_{j}^{(n)}\right]   &=J_{ij}^{(m+n+1)}\,, \\
		\left[J_{ij}^{(m)}, G_k^{(n)}\right]  & =2 \delta_{k[j} G_{i]}^{(m+n)}\,, &
		\left[J_{ij}^{(m)}, J_{k l}^{(n)}\right]  &=4 \delta_{[i[k} J_{l] j]}^{(m+n)}\,,\\
		\left[J_{ij}^{(m)}, Z_{k l}^{(n)}\right]  & =4 \delta_{[i[k} Z_{l] j]}^{(m+n)}\,, &
		\left[J_{ij}^{(m)}, Z_k^{(n)}\right]   &=2 \delta_{k[j} Z_{i]}^{(m+n)}\,, \\
		\left[Z_{ij}^{(m)}, G_k^{(n)}\right]  & =2 \delta_{k[j} Z_{i]}^{(m+n)}\,, &
		\left[G_{i}^{(m)}, Z_{j}^{(n)}\right]   &=Z_{ij}^{(m+n+1)}\,, \\
		\left[P_{i}^{(m)}, H^{(n)}\right]  & = -Z_{i}^{(m+n)} \,, &
		\left[P_{i}^{(m)}, P_{j}^{(n)}\right]   &=Z_{ij}^{(m+n)}\,.&
	\end{align}
\end{subequations}	
The dimensions of the various generators are:
$ [H^{(0)}]=T L^{-2} $ dual to $ s $, $ [P^{(0)}_i] =L^{-1}$, $ [G_i^{(0)}]=T L^{-1} $, dual to Carroll boosts, $ \beta^i_{\text{Carroll}} $, $ [J_{ij}^{(0)}]=1 $, $ [Z_i^{(0)}]= L^{-3} T $ and $ Z_{ij}^{(0)}=L^{-2} $. 

We define local coordinates $ x^i_{(m)}, s_{(m)}, \theta^{i}_{(m)}  $ and $ \phi^{ij}_{(m)} $ dual to $ P_i^{(m)} ,H^{(m)},Z_{i}^{(m)},Z_{ij}^{(m)}$ and group them into the collective coordinates:
\begin{align}
	X^i&=\sum_{m=0}^\infty C^{-2m}x^i_{(m)}, & S&=\sum_{m=0}^{\infty}C^{-2m}s_{(m)} ,& \Theta^{i} &= \sum_{m=0}^\infty C^{-2m-1}\theta^{i}_{(m)}, &  \Phi^{ij} &= \sum_{m=0}^\infty C^{-2m}\phi^{ij}_{(m)},\nn\\
	&& F_{0i}&=\sum_m C^{-2m+1}f_{0i}^{(m)},& F_{ij}&=\sum_m C^{-2m}f_{ij}^{(m)},& 
\end{align}
such that $ [\Theta^i]=L^2 $ and the field $ f_{0i}^{(m)} $ has dimensions of $\tilde{F}^{si}/ C^{2m}$ and $ f_{ij}^{(m)} $ has dimensions  $F^{ij}/C^{2m}$.

{\allowdisplaybreaks
	Expanding the action
	\begin{equation}
		\label{eq:CT}
		S_{\text{tachyon}} =  \int d\tau\left\{-\tilde{M}\left[\sqrt{\dot{\vec{X}}^2-\dot{S}^2/C^2}\right]-\frac 12 F_{\mu\nu}\Omega^{\mu\nu}\right\} 
	\end{equation}
	of a Carroll tachyon in a constant electro-magnetic background, we get
	\begin{subequations}
		\label{key}
		\begin{align}
			S_{(0)}&=-\int d\tau\left\{\tilde{M} \sqrt{\dot{\vec{x}}^2_{(0)}}-f_{0i}^{(0)}\left(\dot{\theta}^{i}_{(0)}+\frac 1{2} \left[\dot{s}_{(0)}x^i_{(0)}-\dot{x}^i_{(0)}s_{(0)}\right]\right)-\frac 12 f_{ij}^{(0)}\left(\dot{\phi}^{ij}_{(0)}+\dot{x}^{[i}_{(0)}{x}^{j]}_{(0)}\right)\right\}\,,\\
			S_{(1)}&=-\frac{1}{C^2}\int d\tau\Bigg\{\tilde{M} \Bigg[-\frac{\dot{s}^2_{(0)}}{2\sqrt{\dot{x}_{(0)}^2}}+\frac{\dot{\vec{x}}_{(0)}\cdot \dot{\vec{x}}_{(1)}}{\sqrt{\dot{x}_{(0)}^2}} \Bigg] -f_{0i}^{(1)}\left(\dot{\theta}^{i}_{(0)}+\frac 1{2} \left[\dot{s}_{(0)}x^i_{(0)}-\dot{x}^i_{(0)}s_{(0)}\right]\right) \nn\\*
			&\hspace{20mm}-\frac 12 f_{ij}^{(1)}\left(\dot{\phi}^{ij}_{(0)}+\dot{x}^{[i}_{(0)}{x}^{j]}_{(0)}\right)\nn\\*
			&\hspace{20mm}-f_{0i}^{(0)}\left(\dot{\theta}^{i}_{(1)}+\frac 1{2} \left[\dot{s}_{(1)}x^i_{(0)}+\dot{s}_{(0)}x^i_{(1)}-\dot{x}^i_{(1)}s_{(0)}-\dot{x}^i_{(0)}s_{(1)}\right]\right) \,,\nn\\
			&\hspace{20mm}-\frac 12 f_{ij}^{(0)}\left(\dot{\phi}^{ij}_{(1)}+\dot{x}^{[i}_{(1)}{x}^{j]}_{(0)}+\dot{x}^{[i}_{(0)}{x}^{j]}_{(1)}\right)\Bigg\}\\
			S_{(2)}&=-\frac{1}{C^4}\int d\tau \Bigg\{\tilde{M}\left[\frac{\dot{s}^4_{(0)}}{8\sqrt{\dot{\vec{x}}^2_{(0)}}\dot{\vec{x}}_{(0)}^2}-\frac{\left(\dot{\vec{x}}_{(0)}{\cdot} \dot{\vec{x}}_{(1)}\right)^2}{2\sqrt{\dot{\vec{x}}^2_{(0)}}\dot{\vec{x}}_{(0)}^2}+\frac{\dot{s}_{(0)}^2\dot{\vec{x}}_{(0)}{\cdot} \dot{\vec{x}}_{(1)}}{2\sqrt{\dot{\vec{x}}^2_{(0)}}\dot{\vec{x}}_{(0)}^2} 
			+\frac{ 2\dot{s}_{(0)}\dot{s}_{(1)} {-}\dot{\vec{x}}_{(1)}^2 {-}2\dot{\vec{x}}_{(0)}{\cdot}\dot{\vec{x}}_{(2)} }{2\sqrt{\dot{\vec{x}}_{(0)}^2}}
			\right]\nn\\*
			&\hspace{20mm}-f_{0i}^{(2)}\left(\dot{\theta}^{i}_{(0)}+\frac 1{2} \left[\dot{s}_{(0)}x^i_{(0)}-\dot{x}^i_{(0)}s_{(0)}\right]\right)-\frac 12 f_{ij}^{(2)}\left(\dot{\phi}^{ij}_{(0)}+\dot{x}^{[i}_{(0)}{x}^{j]}_{(0)}\right)+\dots\Bigg\}
		\end{align}
	\end{subequations}
}

Let us also consider the equations of motion. The ones implied by $ S_{(0)} $ are:
\begin{subequations}
	\begin{align}
		\delta s_{(0)}:\hspace{18mm}f_{0i}^{(0)}\dot{x}_{(0)i}&=0\,,\\
		\delta x^i_{(0)}:\hspace{5.5mm} \tilde{M}\frac{d}{d\tau }\left[\frac{\dot{x}^i_{(0)}}{\sqrt{\dot{\vec{x}}^2_{(0)}}}\right]{}&=-f_{0i}^{(0)}\dot{s}_{(0)}+f_{ij}^{(0)}\dot{x}^j_{(0)}\,,\\
		\delta \theta^i_{(0)}:\hspace{20mm} \frac{d}{d\tau} f_{0i}^{(0)}{}&=0\,,\\
		\delta \phi^{ij}_{(0)}:\hspace{20mm} \frac{d}{d\tau}f_{ij}^{(0)} {}&=0 \,.
	\end{align}
\end{subequations}

The (new) dynamical equations implied by $ S_{(1)} $ are:
\begin{subequations}
\begin{align}
	\delta x^i_{(0)}&:& \tilde{M}\frac{d}{d\tau }\left[\frac{\dot{x}^i_{(0)}\dot{s}_{(0)}^2}{2\sqrt{\dot{\vec{x}}^2_{(0)}}\dot{\vec{x}}^2_{(0)}}{+}\frac{\dot{x}^i_{(1)}}{\sqrt{\dot{\vec{x}}^2_{(0)}}}{-}\frac{\dot{\vec{x}}_{(0)}\cdot \dot{\vec{x}}_{(1)}\dot{x}^i_{(0)}}{\dot{\vec{x}}^2_{(0)}\sqrt{\dot{\vec{x}}^2_{(0)}}}\right]&=-f_{0i}^{(1)}\dot{s}_{(0)}-f_{0i}^{(0)}\dot{s}_{(1)}+f_{ij}^{(1)}\dot{x}^j_{(0)}+f_{ij}^{(0)}\dot{x}^j_{(1)}\,,\\
	\delta s_{(0)}&:& \tilde{M}\frac{d}{d\tau }\left[\frac{\dot{s}_{(0)}}{\sqrt{\dot{\vec{x}}^2_{(0)}}}\right]&=-f^{(1)}_{0i}\dot{x}^i_{(0)}-f^{(0)}_{0i}\dot{x}^i_{(1)}\,, \\
	\delta \theta^{i}_{(0)}&:& \frac{d}{d\tau} f_{0i}^{(1)}&=0 \,,\\
	\delta \phi^{ij}_{(0)}&:& \frac{d}{d\tau}f_{ij}^{(1)}&=0 \,.
\end{align}
\end{subequations}

We see that they agree with \eqref{eq:covC2}. Under a general element of the algebra
\begin{align}
	\sum_{n=0}^\infty\left( \epsilon_{(n)}H^{(n)}+\epsilon^i_{(n)}P_i^{(n)}+\beta^i_{(n)}G_i^{(n)}+\varepsilon^i_{(n)}Z_i^{(n)}+\varepsilon^{ij}_{(n)}Z_{ij}^{(n)}\right)
\end{align}
the coordinates, dual to the generators defined in \eqref{def:ecGen}, transform as:
\begin{subequations}
	\begin{align}
		\delta s_{(m)}&=\epsilon_{(m)}+\sum_{n=0}^{m} \beta^i_{(m-n)}x^{(n)}_i,\\
		\delta x^i_{(m)}&=\epsilon^i_{(m)}+\sum_{n=0}^{m-1}\beta^i_{(m-n-1)}s_{(n)} ,\\
		\delta \theta^i_{(m)}&=\varepsilon^i_{(m)}+\sum_{n=0}^{m} \frac 12 \epsilon^i_{(m-n)}s_{(n)}-\frac12 \epsilon_{(m-n)}x^i_{(n)}-2\beta^k_{(m-n)}\phi^{i}_{k(n)}, \\
		\delta \phi^{ij}_{(m)}&=\varepsilon^{ij}_{(m)}+\sum_n^{m} \epsilon^{[i}_{(m-n)}x^{j]}_{(n)}-2\beta^{[i}_{(m-n-1)}\theta^{j]}_{(n)}
		\label{transmag2c}.
	\end{align}
\end{subequations}
For invariance of the action we also demand that the fields $ f_{0i}^{(m)} $ and $ f_{ij}^{(m)} $ transform under boosts as:
\begin{subequations}
	\begin{align}
		\delta f_{0i}^{(m)}&=-\sum_{n=0}^{m-1} \beta_{(m-n-1)}^{j}f_{ij}^{(n)}, \\
		\delta f_{ij}^{(m)}&=-2 \sum_{n=0}^m\beta_{[i}^{(m-n)}f_{0|j]}^{(n)}\,.
	\end{align}
\end{subequations}
After  raising/lowering indices, and identifying $ f^{0i}_{(m)}=\tilde{F}^{si}_{(m)} \,, f^{ij}_{(m)}=F^{ij}_{(m)}$, we see that the results from the bottom-up approach are successfully reproduced by the top-down analysis.

\subsubsection{Magnetic Carroll Maxwell}

In the magnetic limit we perform the contraction of the Maxwell algebra according to:
\begin{align} 
	\tilde{M}_{i j} & =M_{i j}, & \tilde{G}_i & =\frac{1}{\omega} M_{i 0} \,,\nn\\ 
	\tilde{H} & =\frac 1 \omega  P_0, & \tilde{P}_i & = P_i \,,\\ 
	\tilde{Z}_{i j} & = Z_{i j}, & \tilde{Z}_i & =\omega Z_{0 i} \,.\nn
\end{align}
The commutation relations after sending $ \omega\rightarrow \infty $ become
\begin{subequations}
\begin{align} 
	{\left[\tilde{G}_i, \tilde{P}_j\right] } & =\delta_{ij}\tilde{H}, & {\left[\tilde{M}_{i j}, \tilde{P}_k\right] } & =2 \delta_{k[j} \tilde{P}_{i]}, & {\left[\tilde{G}_i, \tilde{Z}_j\right] } & =\tilde{Z}_{ij}, \\ 
	{\left[H, \tilde{G}_i\right] } & =0, & {\left[\tilde{M}_{i j}, \tilde{G}_k\right] } & =2 \delta_{k[j} \tilde{G}_{i]}, &  \left[\tilde{P}_i, \tilde{P}_j\right]&=\tilde{Z}_{ij},  \\ 
	{\left[\tilde{H}, \tilde{P}_i\right] } & =0, & {\left[\tilde{M}_{i j}, \tilde{Z}_k\right] } & =2 \delta_{k[j} \tilde{Z}_{i]}, & {\left[\tilde{G}_k, \tilde{Z}_{i j}\right] } & =0, \\
	{\left[\tilde{G}_i, \tilde{G}_j\right] } & =0, & {\left[\tilde{M}_{i j}, \tilde{Z}_{k l}\right] } & =-4 \delta_{[i[l} \tilde{Z}_{k] j]} . & & 
\end{align}
\end{subequations}
For the expansion of the magnetic Carroll, we consider the grading
\begin{align}
	V_0=\{J_{ij}\} \,, && V_1=\{G_i,P_i,Z_i\} \,,&& V_2=\{H,Z_{ij}\}
\end{align}
and the usual semigroups $ S_0^{(\infty)},S_1^{(\infty)},S_2^{(\infty)} $, such that we have the following generators:
\begin{subequations}
\begin{align}
	J^{(m)}_{ij}={M}_{ij}\otimes \lambda_{2m},&&H^{(m)}={P}_0\otimes \lambda_{2m+2}, && 	Z^{(m)}_{ij}={Z}_{ij}\otimes \lambda_{2m+2},\\
	G^{(m)}_{i}={M}_{0i}\otimes \lambda_{2m+1},&&P^{(m)}_i={P}_i\otimes \lambda_{2m+1}, && 	Z^{(m)}_{i}={Z}_{0i}\otimes \lambda_{2m+1}.
\end{align}
\end{subequations}

The commutation relations are:
\begin{subequations}
	\begin{align}
		{\left[G_i^{(m)}, H^{(n)}\right] } & =P_i^{(m+n+1)} &&		{\left[G_i^{(m)}, P_j^{(n)}\right] }  =\delta_{ij} H^{(m+n)}\,, \\
		{\left[J_{ij}^{(m)}, P_k^{(n)}\right] } & =2 \delta_{k[j} P_{i]}^{(m+n)} &&
		{\left[G_i^{(m)}, G_j^{(n)}\right] }  =J_{ij}^{(m+n+1)} \,,\\
		{\left[J_{ij}^{(m)}, G_k^{(n)}\right] } & =2 \delta_{k[j} G_{i]}^{(m+n)} &&
		{\left[J_{ij}^{(m)}, J_{kl}^{(n)}\right] }  =4 \delta_{[i[k} J_{l] j]}^{(m+n)}\,,\\
		{\left[J_{ij}^{(m)}, Z_{kl}^{(n)}\right] } & =4 \delta_{[i[k} Z_{l] j]}^{(m+n)}, &&
		{\left[J_{ij}^{(m)}, Z_k^{(n)}\right] }  =2 \delta_{k[j} Z_{i]}^{(m+n)}, \\
		{\left[Z_{ij}^{(m)}, G_k^{(n)}\right] } & =2 \delta_{k[j} Z_{i]}^{(m+n+1)}, &&
		{\left[G_i^{(m)}, Z_j^{(n)}\right] }  =Z_{ij}^{(m+n)}\,,, \\
		{\left[P_i^{(m)}, H^{(n)}\right] } & =-Z_i^{(m+n+1)}, &&
		{\left[P_i^{(m)}, P_j^{(n)}\right] }  =Z_{ij}^{(m+n)}\,.
	\end{align}
\end{subequations}

Introducing $ \lambda_{2m}=C^{-2m} $ and $ \lambda_{2m+1}=C^{-2m-1} $, and choosing appropriate units, we define the following generators:
\begin{subequations}
\begin{align}
	J^{(m)}_{ij}={M}_{ij}\otimes C^{-2m},&&H^{(m)}={P}_0\otimes C^{-2m-1}, && 	Z^{(m)}_{ij}={Z}_{ij}\otimes C^{-2m},\\
	G^{(m)}_{i}={M}_{0i}\otimes C^{-2m-1},&&P^{(m)}_i={P}_i\otimes C^{-2m}, && 	Z^{(m)}_{i}={Z}_{0i}\otimes C^{-2m+1}.
\end{align}
\end{subequations}

Define the following collective coordinates:
\begin{align}
	X^i&=\sum_{m=0}^\infty C^{-2m}x^i_{(m)}, & S&=\sum_{m=0}^{\infty}C^{-2m}s_{(m)} ,& \Theta^{i} &= \sum_{m=0}^\infty C^{-2m+1}\theta^{i}_{(m)}, &  \Phi^{ij} &= \sum_{m=0}^\infty C^{-2m}\phi^{ij}_{(m)},\nn\\
	&& F_{0i}&=\sum_m C^{-2m-1}f_{0i}^{(m)},& F_{ij}&=\sum_m C^{-2m}f_{ij}^{(m)}.& 
\end{align}
In this way, $f_{0i}^{(m)} $ has dimensions of $ \tilde{F}^{si}/C^{2m+2} $, and $ f_{ij}^{(m)} $ has dimensions of $ {F}^{ij} /C^{2m}$.

The expansion of the action~\eqref{eq:CT} now becomes:
{\allowdisplaybreaks
	\begin{subequations}
		\label{expandedcmag}
		\begin{align}
			S_{(0)}&=-\int d\tau\left\{\tilde{M} \sqrt{\dot{\vec{x}}^2_{(0)}}-f_{0i}^{(0)}\theta^{i}_{(0)}-\frac 12 f_{ij}^{(0)}\left(\dot{\phi}^{ij}_{(0)}+\dot{x}^{[i}_{(0)}{x}^{j]}_{(0)}\right)\right\}\,,\\
			S_{(1)}&=-\frac{1}{C^2}\int d\tau\left\{\tilde{M} \left[-\frac{\dot{s}^2_{(0)}}{2\sqrt{\dot{x}_{(0)}^2}}+\frac{\dot{\vec{x}}_{(0)}\cdot \dot{\vec{x}}_{(1)}}{\sqrt{\dot{x}_{(0)}^2}} \right] -f_{0i}^{(0)}\left(\dot{\theta}^{i}_{(1)}+\frac 1{2} \left[\dot{s}_{(0)}x^i_{(0)}-\dot{x}^i_{(0)}s_{(0)}\right]\right)\right.\nn\\
			&\hspace{20mm}\left.-\frac 12 f_{ij}^{(1)}\left(\dot{\phi}^{ij}_{(0)}+\dot{x}^{[i}_{(0)}{x}^{j]}_{(0)}\right)-\frac 12 f_{ij}^{(0)}\left(\dot{\phi}^{ij}_{(1)}+\dot{x}^{[i}_{(1)}{x}^{j]}_{(0)}+\dot{x}^{[i}_{(0)}{x}^{j]}_{(1)}\right)-f_{0i}^{(1)}\dot{\theta}^{i}_{(0)} \right\}\,,\\
			S_{(2)}&=-\frac{1}{C^4}\int\!\!d\tau \Bigg\{\tilde{M}\Bigg[\frac{\dot{s}^4_{(0)}}{8\sqrt{\dot{\vec{x}}^2_{(0)}}\dot{\vec{x}}_{(0)}^2}-\frac{\left(\dot{\vec{x}}_{(0)}{\cdot} \dot{\vec{x}}_{(1)}\right)^2}{2\sqrt{\dot{\vec{x}}^2_{(0)}}\dot{\vec{x}}_{(0)}^2}+\frac{\dot{s}_{(0)}^2\dot{\vec{x}}_{(0)}\cdot \dot{\vec{x}}_{(1)}}{2\sqrt{\dot{\vec{x}}^2_{(0)}}\dot{\vec{x}}_{(0)}^2} 
			+\frac{ 2\dot{s}_{(0)}\dot{s}_{(1)}{-}\dot{\vec{x}}_{(1)}^2{-}2\dot{\vec{x}}_{(0)}\cdot\dot{\vec{x}}_{(2)} }{2\sqrt{\dot{\vec{x}}_{(0)}^2}}
			\Bigg]\nn\\*
			&\hspace{30mm} -f_{0i}^{(2)}\left(\dot{\theta}^{i}_{(0)}+\frac 1{2} \left[\dot{s}_{(1)}x^i_{(0)}+\dot{s}_{(0)}x^i_{(1)}-\dot{x}^i_{(1)}s_{(0)}-\dot{x}^i_{(0)}s_{(1)}\right]\right)\nn\\*
			&\hspace{30mm}-\frac 12 f_{ij}^{(2)}\left(\dot{\phi}^{ij}_{(0)}+\dot{x}^{[i}_{(0)}{x}^{j]}_{(0)}\right)+\dots\Bigg\}\,.
		\end{align}
	\end{subequations}
	}

The equations of motion for $ S_{(0)} $ are:
\begin{subequations}
	\begin{align}
		\delta x^i_{(0)}:\hspace{5.5mm} \tilde{M}\frac{d}{d\tau }\left[\frac{\dot{x}_{i(0)}}{\sqrt{\dot{\vec{x}}^2_{(0)}}}\right]{}&=f_{ij}^{(0)}\dot{x}^j_{(0)}\,,\\
		\delta \theta^i_{(0)}:\hspace{20mm} \frac{d}{d\tau} f_{0i}^{(0)}{}&=0\,,\\
		\delta \phi^{ij}_{(0)}:\hspace{20mm} \frac{d}{d\tau}f_{ij}^{(0)} {}&=0 \,.
	\end{align}
\end{subequations}

The (new) dynamical equations implied by $ S_{(1)} $ are:
\begin{subequations}
\begin{align}
	\delta x^i_{(0)}&:& \tilde{M}\frac{d}{d\tau }\left[\frac{\dot{x}_{i(0)}\dot{s}_{(0)}^2}{2\sqrt{\dot{\vec{x}}^2_{(0)}}\dot{\vec{x}}^2_{(0)}}+\frac{\dot{x}_{i(1)}}{\sqrt{\dot{\vec{x}}^2_{(0)}}}-\frac{\dot{\vec{x}}_{(0)}\cdot \dot{\vec{x}}_{(1)}\dot{x}_{i(0)}}{\dot{\vec{x}}^2_{(0)}\sqrt{\dot{\vec{x}}^2_{(0)}}}\right]&=-f_{0i}^{(0)}\dot{s}_{(0)}+f_{ij}^{(1)}\dot{x}^j_{(0)}+f_{ij}^{(0)}\dot{x}^j_{(1)}\,,\\
	\delta s_{(0)}&:& \tilde{M}\frac{d}{d\tau }\left[\frac{\dot{s}_{(0)}}{\sqrt{\dot{\vec{x}}^2_{(0)}}}\right]&=-f^{(0)}_{0i}\dot{x}^i_{(0)}\,, \\
	\delta \theta^{i}_{(0)}&:& \frac{d}{d\tau} f_{0i}^{(1)}&=0 \,,\\
	\delta \phi^{ij}_{(0)}&:& \frac{d}{d\tau}f_{ij}^{(1)}&=0 \,.
\end{align}
\end{subequations}
Under the identification $ f^{0i}_{(m)}=\tilde{F}^{si}_{(m+1)} \,, f^{ij}_{(m)}=F^{ij}_{(m)}$ and after the raising/lowering indices we recover \eqref{eq:covC2mag}.

Under a general element of the algebra

\begin{align}
	\sum_{n=0}^\infty\left( \epsilon_{(n)}H^{(n)}+\epsilon^i_{(n)}P_i^{(n)}+\beta^i_{(n)}G_i^{(n)}+\varepsilon^i_{(n)}Z_i^{(n)}+\varepsilon^{ij}_{(n)}Z_{ij}^{(n)}\right)
\end{align}
the coordinates, dual to the generators defined in~\eqref{def:ecGen}, transform as:
\begin{subequations}
	\label{eq:transcelec}
	\begin{align}
		\label{transcelec1}
		\delta s_{(m)}&=\epsilon_{(m)}+\sum_{n=0}^{m} \beta^i_{(m-n)}x^{(n)}_i,\\
		\delta x^i_{(m)}&=\epsilon^i_{(m)}+\sum_{n=0}^{m-1}\beta^i_{(m-n-1)}s_{(n)} ,\\
		\delta \theta^i_{(m)}&=\varepsilon^i_{(m)}+\sum_{n=0}^{m} \frac 12 \epsilon^i_{(m-n)}s_{(n)}-\frac12 \epsilon_{(m-n)}x^i_{(n)}-2\beta^k_{(m-n-1)}\phi^{i}_{k(n)}, \\
		\delta \phi^{ij}_{(m)}&=\varepsilon^{ij}_{(m)}+ \sum_{n=0}^{m} \epsilon^{[i}_{(m-n)}x^{j]}_{(n)}-\beta^{[i}_{(m-n)}\theta^{j]}_{(n)}\,.
	\end{align}
\end{subequations}
Invariance also requires fixing the transformations of $ f_{0i}^{(m)} $ and $ f_{ij}^{(m)} $ under boosts:
\begin{subequations}
	\begin{align}
		\delta f_{0i}^{(m)}&=-\sum_{n=0}^{m} \beta_{(m-n)}^{j}f_{ij}^{(n)}, \\
		\delta f_{ij}^{(m)}&=-2 \sum_{n=0}^{m-1}\beta_{[i}^{(m-n-1)}f_{0|j]}^{(n)}.
	\end{align}
\end{subequations}
This again reproduces the results from the bottom-up approach.

\section{A note on the duality Carroll--Galilei}
\label{sec:duality}

In \cite{Gomis:2022spp} a correspondence was established in the (non-Lorentzian) expansion of a free particle between the massive (massless) Galilei case and the tachyonic (massive) Carroll regime, at the level of the action.
A `duality' between the Carroll and Galilei algebras was also established in~\cite{Barducci:2018wuj,Figueroa_O_Farrill_2023}.
Some subtleties arise when trying to extend this duality in the presence of an electro-magnetic field, both at the level of the (expanded) algebra and at the level of the equations of motion.

At the level of the algebras, the duality between Carroll and Galilei can be seen for instance when studying their contraction from the Poincaré algebra, where a duality $ P_0\leftrightarrow P_i $ arises. Consider the different contractions of the Maxwell algebra  to obtain the Carroll and Galilei electro-magnetic limits presented in Table  \ref{Table:contractions}.

{\allowdisplaybreaks
\begin{table}[ht!]
	\centering
	\begin{tabular}{|c|c|c|}
		\hline
		& Galilei & Carroll\\
		\hline&&\\[-4mm]
		Electric &$ 	\begin{aligned} 
			\tilde{M}_{i j} & =M_{i j}, & \tilde{G}_i & =\frac{1}{\omega} M_{i 0} ,\\
			\tilde{H} & =  P_0, & \tilde{P}_i & =\frac{1}{\omega} P_i ,\\ 
			\tilde{Z}_{i j} & = Z_{i j}, & \tilde{Z}_i & =\frac{1}{\omega} Z_{0 i}.
		\end{aligned} $& $ \begin{aligned}
		\tilde{M}_{i j} & =M_{i j}, & \tilde{G}_i & =\frac{1}{\omega} M_{i 0} \,,\nn\\
		\tilde{H} & = \frac{1}{\omega}P_0, & \tilde{P}_i & = P_i \,,\nn\\ 
		\tilde{Z}_{i j} & =  Z_{i j}, & \tilde{Z}_i & =\frac{1}{\omega} Z_{0 i}.
	\end{aligned} $\\[12mm]
		\hline&&\\[-4mm]
		Magnetic & $ 	\begin{aligned} 
			\tilde{M}_{i j} & =M_{i j}, & \tilde{G}_i & =\frac{1}{\omega} M_{i 0} ,\\
			\tilde{H} & =  P_0, & \tilde{P}_i & =\frac{1}{\omega} P_i ,\\ 
			\tilde{Z}_{i j} & =\frac{1}{\omega^2} Z_{i j}, & \tilde{Z}_i & =\frac{1}{\omega} Z_{0 i}.
		\end{aligned} $ &$ \begin{aligned}
		\tilde{M}_{i j} & =M_{i j}, & \tilde{G}_i & =\frac{1}{\omega} M_{i 0} \,,\nn\\
		\tilde{H} & = \frac{1}{\omega} P_0, & \tilde{P}_i & =  P_i \,,\nn\\ 
		\tilde{Z}_{i j} & =  Z_{i j}, & \tilde{Z}_i & ={\omega} Z_{0 i}.
	\end{aligned} $ \\[12mm]
		\hline
	\end{tabular}
\caption{\textit{Contractions of the Maxwell algebra to the different non-Lorentzian algebras. The conventions are slightly changed from previous sections to match the discussion from \cite{Barducci:2019fjc}.}}
	\label{Table:contractions}
\end{table}}

Inspection of the table shows that there is no obvious duality transformation. The reason turns out to be simply that there is only one longitudinal (time) direction in this discussion and so the there is no generator that arises is naively dual to $Z_{ij}$ since $Z_{00}=0$ by anti-symmetry. 

The situation is different if the relativistic translations would split into $P_\mu= (P_\alpha, P_i)$ as in~ \cite{Barducci:2019fjc}, making use of several longitudinal directions $\alpha$ as would be the case for branes. Then there would be an element $[P_\alpha,P_\beta]=Z_{\alpha\beta} $ in the algebra that could be considered as the dual of $Z_{ij}$ under the exchange of longitudinal and transverse directions.\footnote{The different contractions of the Maxwell algebra in the case of more than one longitudinal direction give eight inequivalent Carroll and Galilei algebras, where the duality between them is explicit and unbroken. They are divided into three groups: of \textit{electric} type, of \textit{magnetic} type and of
	\textit{pulse} type.} If one allows for more longitudinal components, the decomposition of the Maxwell generators is  $ Z_{\mu\nu} =( Z_{ij}, Z_{\alpha i}, Z_{\alpha\beta})$
and we can map
$ P_\alpha\leftrightarrow P_i $, $Z_{\alpha i} \leftrightarrow Z_{\alpha i}$  and $ Z_{ij}\leftrightarrow Z_{\alpha\beta} $. The electro-magnetic type of the contraction gets switched and the pattern is not obvious, as there are multiple electric/magnetic-like contractions when we allow for multiple longitudinal directions. In particular, the magnetic Galilei regime gets mapped via this duality to the electric Carroll regime.

One can check that duality in the case of the expanded algebras and their commutation relations is  not present  at the level of the equations of motion in absence of generators $ P_\alpha $ and $ Z_{\alpha\beta} $ either, where the duality would come from the coordinate of the generators of the algebra, $ x^0\leftrightarrow  \vec{x},\; F^{ij}\leftrightarrow F^{\alpha\beta}$.

The duality between Carroll and Galilei regimes in the presence of an electro-magnetic field, or from an algebra point of view, in the case of their Maxwell extensions, can still be traced back to the interchange of transverse and longitudinal directions. Unfortunately, the antisymmetry of the Maxwell generators $ Z_{ab} $ (or alternatively of the electro-magnetic tensor $F^{ab}$) and the presence of only one longitudinal direction makes this symmetry degenerate.

	\section{Conclusion}
	
In this paper we have investigated several
non-Lorentzian expansions of the Lorentz force for a particle in a constant electro-magnetic field: a Galilean expansion in the case of a massive point particle and a Carrollian one in the case of a tachyonic point particle. In both cases, there are also different limits depending on the relative strength of the electric and magnetic fields: the electric, magnetic and pulse limit.

Expanding the position and time coordinates in powers of $ 1/c^2 $ (or $1/C^2$), as well as a similar expansion for the electric and magnetic field components, we have obtained a series of equations of motion, which can be thought as a series of relativistic corrections to the non-Lorentzian limits.

We also showed that this analysis agrees with a top-down approach based on a Lie algebra expansion of the underlying symmetry algebras.
We obtained the algebra of symmetries of the expanded equations, that turned out to be infinite-dimensional algebras which appear as certain infinite-dimensional expansions using the semigroup expansion method of the Maxwell algebra. We have also shown how to obtain the very same equations of motion from considering non-linear realisations of the infinite-dimensional algebras.

We have also discussed the possible extension of the known duality between the Carroll and Galilei limits in the presence of an electro-magnetic field, via their algebra of symmetries.

Our analysis was restricted to constant electro-magnetic fields and it would be interesting to extend it to varying fields. One possible starting point would be the extension of Maxwell algebras considered in~\cite{Bonanos:2008ez,Gomis:2017cmt} where one would need to consider first the analogues of the various non-Lorentzian limits appearing in the present paper. 

An interesting open question is whether a similar construction can be carried out in the case of a test particle in a fixed gravitational background. For this one would need to study the non-Lorentzian limits of the geodesic equation with an appropriate notion of constancy of the gravitational background. The same issue arises for a coloured particle coupled to a Yang--Mills background~\cite{Wong:1970fu}.

\subsubsection*{Acknowledgements}
We are grateful to Luca Ciambelli, Jaume Gomis, Sabrina Pasterski and Jorge Russo for interesting discussions.
AK gratefully acknowledges the generous hospitality of the Universitat de Barcelona through the Maria de Maeztu programme where this work was initiated.
JG acknowledges the hospitality and support of the Max Planck  Institute for Gravitational Physics in Golm and the Perimeter Institute in Waterloo where this work has been further elaborated.
The research of JG was supported in part by
PID2019-105614GB-C21 and by the State Agency for Research of the Spanish Ministry of Science and Innovation 
through the Unit of Excellence Maria de Maeztu 2020-2023 award to the Institute of Cosmos Sciences (CEX2019-000918-M).

\appendix

	\section{\texorpdfstring{Expanding in $1/c$}{Expanding in 1/c}}
	\label{app:1c}
	
	In this appendix, we consider the possibility of expanding all quantities in terms of powers of $1/c$ rather than $1/c^2$ as done in the main part of the paper in sections~\ref{sec:NRLor} and~\ref{sec:topdown}. This is relevant for including also the pulse limit~\cite{Barducci:2019fjc,Gomis:2019fdh}. We will focus only on the expansion in the Galilei limit, as the Carroll case then is straight-forward.
	
	We write the expansion of the variables in powers of $ 1/c $ instead of $ 1/c^2 $ as
	\begin{align}
		t=t_{(0)}+\frac{1}{c}t_{(1/2)}+\frac 1 {c^2}t_{(1)}+\dots, && x^i=x^i_{(0)}+\frac 1 {c}x^i_{(1/2)}+\frac{1}{c^2}x^i_{(1)}+\dots\nn\\
		\tilde{F}^{ti}=\tilde{F}^{ti}_{(0)}+\frac{1}{c}\tilde{F}^{ti}_{(1/2)}+\frac 1 {c^2}\tilde{F}^{ti}_{(1)}+\dots, && {F}^{ij}={F}^{ij}_{(0)}+\frac 1 {c}{F}^{ij}_{(1/2)}+\frac{1}{c^2}{F}^{ij}_{(1)}+\dots
	\end{align}
	
	We then obtain the following equations for the temporal component of the Lorentz equation~\eqref{eq:LorT} at lowest orders in the expansion:
	\begin{subequations}
		\label{1/cTempLorentz}
		\begin{align}
			m \frac{d}{d\tau} \left[ \frac{\dot{\vec{x}}_{(0)}^2}{2\dot{t}_{(0)}^2} \right] &= \tilde{F}^{ti}_{(0)} \dot{x}_{(0) i}\,,\\
			m  \frac{d}{d\tau}\left[ \frac{  \dot{\vec{x}}_{(0)}\cdot \dot{\vec{x}}_{(1/2)}}{\dot{t}_{(0)}^2} - \frac{\dot{t}_{(1/2)}\dot{\vec{x}}_{(0)}^2 }{\dot{t}_{(0)}^3}
			\right] &=   \tilde{F}^{ti}_{(0)} \dot{x}_{(1/2) i} + \tilde{F}^{ti}_{(1/2)} \dot{x}_{(0) i} 
		\end{align}
		and
		\begin{align}
			&\quad	m  \frac{d}{d\tau}\left[\frac 38 \frac{  \dot{\vec{x}}_{(0)}^4}{\dot{t}_{(0)}^4}+\frac 32 \frac{\dot{t}_{(1/2)}^2\dot{\vec{x}}_{(0)}^2}{\dot{t}_{(0)}^4}-\frac{\dot{t}_{(1)}\dot{\vec{x}}_{(0)}^2}{\dot{t}_{(0)}^3}-2\frac{\dot{t}_{(1/2)}\dot{\vec{x}}_{(0)}\cdot\dot{\vec{x}}_{(1/2)}}{\dot{t}_{(0)}^3}+\frac{\dot{\vec{x}}_{(1/2)}^2}{2\dot{t}_{(0)}^2}+\frac{\dot{\vec{x}}_{(0)}\cdot \dot{\vec{x}}_{(1)}}{\dot{t}_{(0)}^2}
			\right]\nn\\
			&=   \tilde{F}^{ti}_{(0)} \dot{x}_{(1) i} + \tilde{F}^{ti}_{(1)} \dot{x}_{(0) i} + \tilde{F}^{ti}_{(1/2)} \dot{x}_{(1/2) i}  \,.
		\end{align}
	\end{subequations}

	For the spatial component of the Lorentz force~\eqref{eq:LorS} we get
	\begin{subequations}
		\label{1/cSpatialLorentz}
		\begin{align}
			m \frac{d}{d\tau} \left[  \frac{\dot{x}_{(0)}^i}{\dot{t}_{(0)}} \right] &= - \tilde{F}^{it}_{(0)}  \dot{t}_{(0)} + F^{ij}_{(0)} \dot{x}_{(0)j}\\
			m  \frac{d}{d\tau}\left[ 	- \frac{\dot{t}_{(1/2)} \dot{x}_{(0)}^i}{\dot{t}_{(0)}^2} + \frac{\dot{x}_{(1/2)}^i}{\dot{t}_{(0)}}   \right] 
			&=
			- \tilde{F}^{it}_{(0)}\dot{t}_{(1/2)}+  F^{ij}_{(0)}  \dot{x}_{(1/2)j} \nn\\ 
			&\quad - \tilde{F}^{it}_{(1/2)}\dot{t}_{(0)}+  F^{ij}_{(1/2)}  \dot{x}_{(0)j} 
		\end{align}
		as well as
		\begin{align}
			&\quad 	m  \frac{d}{d\tau}\left[  \frac12 \frac{\dot{\vec{x}}_{(0)}^2 \dot{x}_{(0)}^i }{ \dot{t}_{(0)}^3} - \frac{\dot{t}_{(1)}\dot{x}^i_{(0)} }{\dot{t}_{(0)}^2} - \frac{\dot{t}_{(1/2)}\dot{x}^i_{(1/2)}}{\dot{t}^2_{(0)}}+\frac{\dot{x}^i_{(1)}}{\dot{t}_{(0)}}+\frac{\dot{t}_{(1/2)}^2\dot{x}^i_{(0)}}{\dot{t}_{(0)}^3}   \right] 	\nn\\
			&=	- \tilde{F}^{it}_{(0)}\dot{t}_{(1)}- \tilde{F}^{it}_{(1/2)}\dot{t}_{(1/2)}	- \tilde{F}^{it}_{(1)}\dot{t}_{(0)}+  F^{ij}_{(1)}  \dot{x}_{(0)j}+F^{ij}_{(1/2)}  \dot{x}_{(1/2)j}+F^{ij}_{(0)}  \dot{x}_{(1	)j}  
		\end{align}
	\end{subequations}

In this context the magnetic, electric and pulse limits are obtained as follows:
	\begin{enumerate}
		\item Magnetic: By keeping $ F^{ij}_{(0)}\neq0 $. The equations of motion are formally the same as (\ref{1/cTempLorentz}, ~\ref{1/cSpatialLorentz}).
		\item Electric: Obtained by setting $ F^{ij}_{(0)}=F^{ij}_{(1/2)}=0 $ and keeping $ \tilde{F}^{ti}_{(0)}\neq0 $.
		The equations of motion are~\eqref{1/cTempLorentz} for the temporal component and for the space components:
		\begin{subequations}
			\label{LorSElec}
			\begin{align}
				m \frac{d}{d\tau} \left[  \frac{\dot{x}_{(0)}^i}{\dot{t}_{(0)}} \right] &= - \tilde{F}^{it}_{(0)}  \dot{t}_{(0)} + F^{ij}_{(0)} \dot{x}_{(0)j}\\
				m  \frac{d}{d\tau}\left[ 	- \frac{\dot{t}_{(1/2)} \dot{x}_{(0)}^i}{\dot{t}_{(0)}^2} + \frac{\dot{x}_{(1/2)}^i}{\dot{t}_{(0)}}   \right] 
				&=
				- \tilde{F}^{it}_{(0)}\dot{t}_{(1/2)}- \tilde{F}^{it}_{(1/2)}\dot{t}_{(0)}\\
				m  \frac{d}{d\tau}\left[  \frac12 \frac{\dot{\vec{x}}_{(0)}^2 \dot{x}_{(0)}^i }{ \dot{t}_{(0)}^3} - \frac{\dot{t}_{(1)}\dot{x}^i_{(0)} }{\dot{t}_{(0)}^2} -\right.{}&\left. \frac{\dot{t}_{(1/2)}\dot{x}^i_{(1/2)}}{\dot{t}^2_{(0)}}+\frac{\dot{x}^i_{(1)}}{\dot{t}_{(0)}}+\frac{\dot{t}_{(1/2)}^2\dot{x}^i_{(0)}}{\dot{t}_{(0)}^3}   \right] 	\nn	\\&=	- \tilde{F}^{it}_{(0)}\dot{t}_{(1)}- \tilde{F}^{it}_{(1/2)}\dot{t}_{(1/2)}
				- \tilde{F}^{it}_{(1)}\dot{t}_{(0)}+  F^{ij}_{(1)}  \dot{x}_{(0)j} \\
				\dots\nn &
			\end{align}
		\end{subequations}
		\item Pulse: This is obtained by setting $ F^{ij}_{(0)}=0 $ and keeping $ \tilde{F}^{ti}_{(0)},F^{ij}_{(1/2)}\neq 0 $.
		The temporal equations are again \eqref{1/cTempLorentz}, while the space component are:
		\begin{subequations}
			\begin{align}
				m \frac{d}{d\tau} \left[  \frac{\dot{x}_{(0)}^i}{\dot{t}_{(0)}} \right] &= - \tilde{F}^{it}_{(0)}  \dot{t}_{(0)} + F^{ij}_{(0)} \dot{x}_{(0)j}\\
				m  \frac{d}{d\tau}\left[ 	- \frac{\dot{t}_{(1/2)} \dot{x}_{(0)}^i}{\dot{t}_{(0)}^2} + \frac{\dot{x}_{(1/2)}^i}{\dot{t}_{(0)}}   \right] 
				&=
				- \tilde{F}^{it}_{(0)}\dot{t}_{(1/2)}- \tilde{F}^{it}_{(1/2)}\dot{t}_{(0)}+F^{ij}_{(1/2)}\dot{x}_{(0)j}\\
				m  \frac{d}{d\tau}\left[  \frac12 \frac{\dot{\vec{x}}_{(0)}^2 \dot{x}_{(0)}^i }{ \dot{t}_{(0)}^3} - \frac{\dot{t}_{(1)}\dot{x}^i_{(0)} }{\dot{t}_{(0)}^2} -\right.{}&\left. \frac{\dot{t}_{(1/2)}\dot{x}^i_{(1/2)}}{\dot{t}^2_{(0)}}+\frac{\dot{x}^i_{(1)}}{\dot{t}_{(0)}}+\frac{\dot{t}_{(1/2)}^2\dot{x}^i_{(0)}}{\dot{t}_{(0)}^3}   \right] 	\nn	\\&=	- \tilde{F}^{it}_{(0)}\dot{t}_{(1)}{-} \tilde{F}^{it}_{(1/2)}\dot{t}_{(1/2)}
				{-} \tilde{F}^{it}_{(1)}\dot{t}_{(0)}{+}  F^{ij}_{(1)}  \dot{x}_{(0)j} {+}F^{ij}_{(1/2)}  \dot{x}_{(1/2)j}  \\
				\dots\nn &
			\end{align}
		\end{subequations}
	\end{enumerate}
It can be checked that all the equations above are invariant under the Lorentz transformations expanded in $ 1/c $ when taking the appropriate limit.
	
	Even though the equations of motion in the electric/magnetic limit in the $1/c$-expansion are different from those in the $1/c^2$-expansion, they realise the same algebra of symmetries as in the $1/c^2$-expansion, namely \eqref{eq:algelec} and \eqref{eq:algmag} respectively. In fact, equations (\ref{1/cTempLorentz}--\ref{LorSElec}) can be obtained from the Lie algebra approach, by setting the expansion parameter $ \lambda=1/\sqrt{c} $, so that $ \lambda_{2m}=c^{-m}$.
	
	Another interesting observation, already mentioned in~\cite{Elbistan:2022plu}, is that the $ 1/c^2 $ expansion can be obtained from the $ 1/c $ expansion via a reshuffling of the terms in the $ 1/c $ expansion.
	In the free case, both the $ 1/c $ and $ 1/c^2 $ expansion yield the same equations of motion. In the presence of an electro-magnetic field, the finer $ 1/c $ equations allow us not only to detect the case of the pulse limit, but also give corrections with respect to the $ 1/c^2 $ expansion.

One way to check if they are describing the same physics is to use the gauge condition and projection used in \cite{Gomis:2019sqv}:
\begin{equation}		
\label{eq:pc}
\frac 1 {c^{2m}}x^i_{(m)}=x^i\,,\hspace{10mm} \frac 1 {c^{2m}}t_{(m)}=t=\tau\,,
\end{equation}
where $ m\in\frac 12\mathbb{Z}  $.
Here, we have only written a projection for the space-time coordinates since we do not know how to properly define a similar projection for the electro-magnetic field.		
		
Thus, focusing only on the free part of the action without electro-magnetic field, the expanded action in $ 1/c $ is
		\begin{align}
				\label{1/caction}
				S & =\int d\tau \Bigg\{ -mc^2\dot{t}_{(0)}-mc\left[\dot{t}_{(1/2)}\right]-m\left[\dot{t}_{(1)}-\frac{\dot{x}^2_{(0)}}{2\dot{t}^2_{(0)}}\right]\nn\\
				&\hspace{20mm}-m\frac{1}{c}\left[\dot{t}_{(3/2)}+\frac{\dot{t}_{(1/2)}\dot{x}^2_{(0)}}{2\dot{t}_{(0)}^3}-\frac{\dot{x}^i_{(1/2)}\dot{x}_{(0)i}}{\dot{t}^2_{(0)}}\right]+\dots\Bigg\}
		\end{align}
		After the projection the $1/c$-expanded action reads:
\begin{subequations}
\begin{align}
				S_{(0)}&=\int d\tau \left[-mc^2\right]\\
				S_{(1/2)}&=\int d\tau \left[-mc^2\right]\\
				S_{(1)}&=\int d\tau \left[-mc^2+\frac 12 m \dot{\vec{x}}^2\right]\\
				S_{(3/2)}&=\int d\tau \left[-mc^2+\frac 12 m \dot{\vec{x}}^2\right]\\
				&\dots\nn
\end{align}
\end{subequations}
We see that the $1/c$ expansion produces the same results as the $1/c^2$ expansion after imposing the projection condition~\eqref{eq:pc}. In particular, all the equations from the $1/c^2$ expansion are simply duplicated.

\section{Free Lie algebras}
	\label{app:free}

	The infinite dimensional algebras $ \mathfrak{M}_\infty $ and $ \mathfrak{E}_\infty $ can also be obtained as particular quotients of the Galilean free algebras. 
	We do not consider the Carroll case here, the extension to Carroll is straight-forward.
	A free Lie algebra with $D=d+1$ generators $ \{P_\mu\} $, is the Lie algebra whose elements are all possible multi-commutators of the generators. The only relations imposed are antisymmetry and the Jacobi identity. The free Lie algebra $\mathfrak{f}$ admits a natural $ \mathbb{N} $-grading
	\begin{align}
		\label{eq:FLAdec}
		\mathfrak{f} = \bigoplus_{\ell=1}^\infty \mathfrak{f}_\ell
	\end{align}
	where each summand $ \mathfrak{f}_\ell $ consists of all possible multi-commutators with $ \ell  $ elements.
	
The recursive relation between each level and the lower ones can be summarised by a generating series identity that reads~\cite{Cederwall:2015oua,Gomis:2018xmo} 
	\begin{align}
		\bigotimes_{\ell=1}^\infty \left[ \bigoplus_{k=0}^\infty (-1)^k t^{k\ell}  \ALT^k \mathfrak{f}_\ell \right] = 1- t \mathfrak{f}_1
	\end{align}
	
	Elements in $\mathfrak{f}_\ell$ can be grouped by representations of the  symmetric group $\mathcal{S}_D$ acting on the elements in a set of multi-commutator of $\ell$ free Lie algebra generators. For this reason we can represent them using Young diagrams with $\ell$ boxes, for example
	\ytableausetup
	{boxsize=1em}\ytableausetup
	{aligntableaux=center}
	\begin{align}
		\label{eq:Fyng}
		\mathfrak{f}_1 \leftrightarrow \ydiagram{1}\,,\quad\quad
		\mathfrak{f}_2 \leftrightarrow \ydiagram{1,1}\,,\quad\quad
		\mathfrak{f}_3 \leftrightarrow \ydiagram{2,1}\quad \text{etc.}
	\end{align}
	The relation of free Lie algebras to Chevalley--Eilenberg cohomology was disccussed in~\cite{Gomis:2017cmt}. Free Lie algebras can be also defined in the superalgebra case, see for example~\cite{Cederwall:2015oua,Gomis:2018xmo}.
	
	Thinking of the elements of $\mathfrak{f}_1$ as  the translation generators of some kinematic algebra, we will also make use of the action of some Lorentz-type of algebra on them. We call the corresponding Lie algebra $\mathfrak{f}_0=\{ M_{\mu\nu} \}$ and then have a graded structure
	\begin{align}
		\mathfrak{f}_0\oplus \mathfrak{f}_1
	\end{align}
	to begin with. The action of $\mathfrak{f}_0$ extends to all $\mathfrak{f}_\ell$ by using the Leibniz property.
	
	\medskip		
	
	We wish to construct free Lie algebras for the electric and magnetic Maxwell algebras in the non-Lorentzian limits. This requires making a choice of starting generators adapted to the non-Lorentzian limit in question.
	
	For the Galilean electric free algebra we then obtain the first levels shown in Table~\ref{tab:GEF}.
	
	\ytableausetup
	{boxsize=0.45em}\ytableausetup
	{aligntableaux=center}
	\begin{table}[ht!]
		\centering
		\scalebox{0.9}{
			\begin{tabular}{c|c|c|c|c|p{3cm}|p{6cm}}
				\centering
				& $\ell=0$& $\ell=1$& $\ell=2$& $\ell=3$& $\ell=4$& $\ell=5$\\
				\hline
				$m=0$ &$M_{ij} $& $ G_i $& $ S_{ij}$& $ Y_{ij,k}$&$ 
				\ydiagram{3,1}\oplus 
				\ydiagram{2,1,1}$ &$ 3\,
				\ydiagram{2,2,1}\oplus 3\, 
				\ydiagram{2,1,1,1}\oplus 
				\ydiagram{3,2}\oplus 3\, 
				\ydiagram{3,1,1}\oplus 
				\ydiagram{4,1} $ \\
				\hline
				$m=1$ & & $H  $&$ P_i$&$ B_{ij} , Z_{i,j} $& $ 
				\ydiagram{3}\oplus 2\,
				\ydiagram{2,1}\oplus 
				\ydiagram{1,1,1} $& $4\,
				\ydiagram{2,2}\oplus 
				\ydiagram{4}\oplus 9 \,
				\ydiagram{2,1,1}\oplus 5 \,
				\ydiagram{3,1}\oplus 3\,
				\ydiagram{1,1,1,1}$\\
				\hline
				$m=2$ & & &$ Z_{ij} $ &$ 
				\ydiagram{1}\oplus 
				\ydiagram{2,1}\oplus 
				\ydiagram{1,1,1} $ &$
				\ydiagram{2}\oplus 2 \,
				\ydiagram{1,1}\oplus 
				\ydiagram{2,2}\oplus 
				\ydiagram{3,1}\oplus 
				\ydiagram{2,2,1}\oplus 
				\ydiagram{2,1,1}\oplus 
				\ydiagram{1,1,1,1}$ & $3\,
				\ydiagram{1,1,1,1,1}\oplus 2 \,
				\ydiagram{3}\oplus 7 \,
				\ydiagram{2,2,1}\oplus 8 \,
				\ydiagram{2,1}\oplus 4 \,
				\ydiagram{3,2}\oplus 6 \,
				\ydiagram{1,1,1}\oplus 7\,
				\ydiagram{2,1,1,1}\oplus 5 \,
				\ydiagram{3,1,1}\oplus
				\ydiagram{4,1}$\\
				\hline
				$ m=3 $ & & & &$ 
				\ydiagram{1,1} $ &$ 
				\ydiagram{1}\oplus 2\,
				\ydiagram{2,1}\oplus 2\,
				\ydiagram{1,1,1} $& $ 7\,
				\ydiagram{1,1,1,1}\oplus 2 \,
				\ydiagram{2}\oplus4\,
				\ydiagram{1,1}\oplus 7 \,
				\ydiagram{2,2}\oplus 12 \,
				\ydiagram{2,1,1}\oplus 5\,
				\ydiagram{3,1}$\\
				\hline
				$ m=4 $ &&&&&$ 
				\ydiagram{1,1}\oplus 
				\ydiagram{2,1,1} $& $ 5\,
				\ydiagram{1,1,1}\oplus
				\ydiagram{1}\oplus 4\,
				\ydiagram{2,1,1,1}\oplus 2 \,
				\ydiagram{3,1,1}\oplus 
				\ydiagram{1,1,1,1,1}\oplus 4\, 
				\ydiagram{2,2,1}\oplus 5 \,
				\ydiagram{2,1}\oplus 
				\ydiagram{3,2}$\\
				\hline
				$ m=5 $ & & & & & & $ 3\,
				\ydiagram{2,1,1}\oplus 
				\ydiagram{1,1,1,1}\oplus 
				\ydiagram{1,1}\oplus
				\ydiagram{2,2}$
			\end{tabular}
		}
		\caption{\label{tab:GEF}\textit{Table with all generators up to level $\ell=5 $ of the Galilean electric free lie algebra. We have used Young tableaux notation to represent $\mathfrak{gl}(D)- $ tensors. A double grading $ (\ell,m) $ was used, following~\cite{Gomis:2019fdh}.}}
	\end{table}
	
	We will be particularly interested in a quotient of the Galilean electric free algebra which we will denote $ \mathfrak{E}_\infty $ and shown in Table~\ref{tab:GEFq}.
	
	\begin{table}[ht!]
		\centering
		\begin{tabular}{c|c|c|c|c|c|c}
			& $\ell=0$& $\ell=1$& $\ell=2$& $\ell=3$& $\ell=4$& $\ell=5$\\
			\hline
			$m=0$ & $J_{ij}^{(0)}$&  $G_i^{(0)}$&$J_{ij}^{(1)}=S_{ij}$& $G_i^{(1)}=Y_i$&$J_{ij}^{(2)}$& $G_i^{(2)}$\\
			$m=1$ & -&$H^{(0)}$&  $P_i^{(0)}$&$H^{(1)}=N=\frac 1 {D-1}\delta^{ij}Z_{i,j}$& $P_i^{(1)}$&$H^{(2)}$\\
			$m=2$ & -& -&$Z_{ij}^{(0)}$& $Z_i^{(0)}=[H,P_i]$&$Z_{ij}^{(1)}$& $Z_i^{(1)}$
		\end{tabular}
		\caption{\label{tab:GEFq}\textit{First few generators of the quotient $ \mathfrak{E}_\infty $ of the Galilean electric free algebra.}}
	\end{table}
		For the Galilean Magnetic free algebra we obtain Table~\ref{tab:GMF}.
	\ytableausetup
	{boxsize=0.4em}\ytableausetup
	{aligntableaux=center}
	\begin{table}[ht!]
		\begin{tabular}{c|c|c|c|c|c|c}
			& $\ell=0$& $\ell=1$& $\ell=2$& $\ell=3$& $\ell=4$& $\ell=5$\\
			\hline
			$m=0$ &$\textcolor{black}{M_{ij}} $& $ G_i $& $ \textcolor{black}{S_{ij}}$& $ Y_{ij,k}$&$
			\textcolor{black}{\ydiagram{3,1}\oplus 
				\ydiagram{2,1,1}}$& $ 3\,\ydiagram{2,2,1}\oplus 3\,\ydiagram{2,1,1,1}\oplus \ydiagram{3,2}\oplus 3\,\ydiagram{3,1,1}\oplus\ydiagram{4,1} $\\
			$m=1$ & & $\textcolor{black}{H}  $&$ P_i$&$ \textcolor{black}{B_{ij} , Z_{i,j}} $& $ 
			\ydiagram{3}\oplus 2\,
			\ydiagram{2,1}\oplus 
			\ydiagram{1,1,1} $& $\textcolor{black}{3\,\ydiagram{1,1,1}\oplus 4\,\ydiagram{2,2}\oplus9\,\ydiagram{2,1,1}\oplus5\,\ydiagram{3,1}\oplus12\,\ydiagram{4}}$\\
			$m=2$ & & &$ $ &$ Z_{i} $ &$\textcolor{black}{
				\ydiagram{2}\oplus  
				\ydiagram{1,1}\oplus Z_{ij}}$ &$6\,\ydiagram{1,1,1}\oplus8\,\ydiagram{2,1}\oplus2\,\ydiagram{3}$\\
			$ m=3 $ & & & & &$ 
			\ydiagram{1} $& $ 4\,\ydiagram{1,1}\oplus2\,\ydiagram{2}$\\
			$	m=4$	& & & & & &  $\ydiagram{1}$
		\end{tabular}
		\caption{\label{tab:GMF}\textit{Table with all generators up to level $ \ell=5  $ of the Galilean magnetic free algebra.}}
	\end{table}
	
 The quotient we will be interested in will be the one from Table~\ref{tab:GMFq2} and we will denote it by $ \mathfrak{M}_\infty $.

	\ytableausetup
	{boxsize=0.5em}\ytableausetup
	{aligntableaux=center}
	\begin{table}[ht!]
		\centering
		\begin{tabular}{c|c|c|c|c|c|c}
			& $\ell=0$& $\ell=1$& $\ell=2$& $\ell=3$& $\ell=4$& $\dots$\\
			\hline
			$m=0$ &$J^{(0)}_{ij} $& $ G^{(0)}_i $& $ J^{(1)}_{ij}=S_{ij}$& $ G^{(1)}_i$&$
			J_{ij}^{(2)}$& $\dots$\\
			$m=1$ & & $H^{(0)}  $&$ P^{(0)}_i$&$ H^{(1)} $& $ 
			P^{(1)} $& $\dots$\\
			$m=2$ & & &$ $ &$ Z^{(0)}_{i} $ &$
			Z^{(0)}_{ij}$ &$\dots$
		\end{tabular}
		\caption{\label{tab:GMFq2}\textit{First few generators of the quotient $\mathfrak{M}_\infty$ of the magnetic free algebra.}}
	\end{table}
	
		Another interesting truncation of both the electric and magnetic Galilean free algebras is $ \mathfrak{G}_\infty $, presented in Table~\ref{tab:GMFq} which was discussed in \cite{Gomis:2019sqv}.
			\ytableausetup
		{boxsize=0.75em}\ytableausetup
		{aligntableaux=center}
		\begin{table}[h!]
			\centering
			\begin{tabular}{c|c|c|c|c|c|c}
				& $\ell=0$& $\ell=1$& $\ell=2$& $\ell=3$& $\ell=4$& $\dots$\\
				\hline
				$m=0$ &$M_{ij} $& $ G_i $& $ S_{ij}$& $ Y_{i}=\delta^{jk}Y_{ij,k}$&$
				\ydiagram{1,1}$& $ \dots $\\
				$m=1$ & & $H  $&$ P_i$&$N=\delta^{ij}Z_{i,j} $& $ 
				\ydiagram{1} $& $\dots$\\
			\end{tabular}
			\caption{\label{tab:GMFq}\textit{Table with the first few generators of the quotient $\mathfrak{G}_\infty$ of the magnetic free algebra.}}
		\end{table}

	
\providecommand{\href}[2]{#2}\begingroup\raggedright\endgroup

\end{document}